\documentclass[aps,prd,showpacs,twocolumn,floatfix,nofootinbib]{revtex4}
\usepackage{epsfig,graphics,color}
\usepackage{graphicx}% Include figure files
\usepackage{dcolumn}% Align table columns on decimal point
\usepackage{bm}% bold math
\usepackage{amsmath}

\newcommand \f {\not\!}

\newcommand \kd  {\delta}
\newcommand \ra  {\rightarrow}

\newcommand \g {\gamma}

\newcommand \e {\epsilon}

\newcommand \x {\cdot}

\newcommand \A {\alpha}
\newcommand \B {\beta}

\newcommand \lc {\langle}
\newcommand \rc {\rangle}
\newcommand \prt {\partial}

\newcommand \dsc {\mbox{Disc}}

\newcommand \bvec{\left( \begin{array}{c} }
\newcommand \evec{\end{array} \right)}
\newcommand \tr {\mbox{{\bf Tr}}}

\newcommand \bea{\begin{eqnarray} }
\newcommand \eea{\end{eqnarray} } 
\newcommand \nn {\nonumber}
\newcommand {\be} {\begin{equation}}
\newcommand {\ee} {\end{equation}}
\newcommand {\epem} {$e^+ e^-$}
\newcommand {\mbx} {\mbox{}}

\newcommand {\ata} {& \times &}
\newcommand {\psibar} {\bar{\psi}}
\begin{document}

\title{Calculating the Jet Quenching Parameter $\hat{q}$ in Lattice Gauge Theory }

\author{Abhijit Majumder}
\affiliation{Department of Physics and Astronomy, Wayne State University, Detroit, MI 48201.}
\affiliation{Department of Physics, Ohio State University, Columbus, OH 43210.}

\date{\today}

\begin{abstract} 
We present a framework where first principles calculations of jet modification may be 
carried out in a non-perturbative thermal environment. As an example of this approach, we compute 
the leading order contribution to the transverse momentum broadening of a high energy (near on-shell) 
quark in a thermal medium. 
This involves a factorization of a non-perturbative operator product from the perturbative process of 
scattering of the quark. An operator product expansion of the non-perturbative operator product is carried out and
related via dispersion relations to the expectation of local operators. 
These local operators are then evaluated in quenched $SU(2)$ lattice gauge theory.

\end{abstract}

\maketitle

\section{Introduction}

As of this time, the Large Hadron Collider (LHC) has completed three successful runs with heavy-ions. 
There is now a wealth of data on the modification of hard jets from the 
Relativistic Heavy-Ion Collider (RHIC)~\cite{Adcox:2004mh,Adams:2005dq} and the LHC~\cite{Schukraft:2011ch,Steinberg:2011qq,Wyslouch:2011zz}. 
With the similarity between the various soft observables between RHIC and LHC the study of 
jets has moved to the forefront of heavy-ion programs at both these colliders.

In the last several years, the science of jet quenching has undergone considerable evolution. There are now four different 
successful jet quenching formalisms based on perturbative QCD (pQCD)~\cite{Armesto:2003jh,Salgado:2003gb,Wiedemann:2000za,Arnold:2002ja,Jeon:2003gi,Qin:2009bk,Gyulassy:2000er,Gyulassy:2000fs,Gyulassy:2001nm,Guo:2000nz,Wang:2001ifa, Majumder:2007hx,Majumder:2007ne,Majumder:2009ge} and a collection of formalisms based on AdS/CFT~ \cite{Gubser:2006bz,Gubser:2009sn,Herzog:2006gh, CasalderreySolana:2007qw}. 
While one would have expected a large disparity between the physical pictures underlying the strong and weak coupling approaches, 
there are actually considerable differences between the various pQCD based approaches~\cite{Majumder:2010qh,Armesto:2011ht}. Besides the differences in the 
description of the perturbative gluon emission process, the description of the medium is quite different in the various approaches: 
In both the Armesto-Salgado-Weidemann (ASW) and the Higher-Twist (HT) approach, one assumes that the transverse momentum 
exchanged in numerous interactions with the medium is soft enough that one may approximate the distribution as a 
Gaussian, and retain only the leading two moments (mean and variance). 
The variance of this Gaussian transverse momentum distribution is often referred to 
as $\hat{q}$. In the Gyulassy-Levai-Vitev (GLV) formalism, the exchanged momentum is assumed to have a considerable hard tail, such 
that it cannot be approximated as a Gaussian broadening. In the Arnold-Moore-Yaffe (AMY) formalism one describes the medium using Hard-Thermal-Loop 
improved perturbation theory~\cite{Braaten:1989kk,Braaten:1989mz}. 

With the exception of the AMY formalism, none of the other pQCD based formalisms can be said to be a first principles calculation. In all cases 
the transport parameter $\hat{q}$ (either averaged or a normalized function of space-time in a fluid dynamical simulation) is a fit parameter 
in the calculation, set by comparison to one data point. Even in the AMY formalism the strong coupling constant $\A_{s}$ is varied to fit one 
data point. Thus, even the AMY formalism is not, strictly speaking, a first principles calculation. The strong coupling approaches, though first 
principles calculations, are not sufficiently sophisticated to address the great variety of jet modification data. The predictions from such calculations 
also seem to be inconsistent with the rising $R_{AA}$ observed at the LHC~\cite{Horowitz:2011wm}.

The goal of the present paper is to 
suggest a setup where a first principles calculation of jet modification can be carried out using a combination of perturbative and non-perturbative 
methods. The perturbative sector will be similar in form with the higher-twist approach in that it will involve a factorization of the perturbative sector describing 
the propagation of hard partons from operator products which will be used to describe the medium. The computation of these operator products in the non-perturbative 
sector will be carried out using finite temperature lattice gauge theory. 
We would point out already at this stage that a completely first principles calculation can never be directly compared with data. It will, 
however, provide constraints on the number, structure and normalization of the various transport coefficients that one routinely 
uses to construct a phenomenological analysis of the data.

The paper is organized as follows: In Sect. II, we describe the set up where calculations can be carried out and in particular we will 
attempt to justify why the current method to identify and estimate jet transport coefficients is the better alternative. 
In Sect. III we will focus on the 
particular process of a hard quark propagating through a medium and set up the formalism for this process. In Sect. IV the various regions of 
phase space will be explored. In Sect. V, dispersion relations that will be used to evaluate the operator products will be set up. In Sect. VI we 
discuss the details of the lattice gauge theory calculation. We conclude in Sect. VII with an outlook for future work.

\section{pQCD processes in a QGP brick}

The notion that jet transport coefficients represent properties of the medium and thus should be calculable in lattice QCD has 
definitely been informally considered for some time now. The most naive approach would be to simply take the 
expression for a given transport coefficient, say $\hat{q}$, as derived in an appropriate effective theory in Ref~\cite{Idilbi:2008vm}, where 
\bea
\hat{q} &=& \frac{ 4\pi^{2} \A_{s} }{N_{c}}  \int \frac{ d y^{-} d^{2} y_{\perp} d^{2} k_{\perp}}{(2\pi)^{3}} e^{i \frac{k_{\perp}^{2} y^{-}}{2 q^{-}} 
-i k_{\perp} \cdot y_{\perp} }  \nn \\
&\times& \left\langle P \left| \tr \left[ t^{a} {F_{\perp}^{a}}^{+ \mu} (y^{-}, y_{\perp})
 t^{b}{ F_{\perp}^{b}}^{+}_{,\mu} \right] \right| P \right\rangle ,
\eea
and attempt to compute this on the lattice (In the equation above ${F_{\perp}}^{\mu \nu}$ is a gauge field strength operator, 
one of whose indices are either $1$ or $2$). This particular form of the transport coefficient is obtained in either covariant 
gauge or light-cone gauge.

The equation above is not manifestly gauge invariant and requires the introduction of Wilson lines. 
At first sight, the path taken by the Wilson lines seems arbitrary. However, following the arguments in Ref.~\cite{GarciaEchevarria:2011md}, 
one obtains four different Wilson lines that need to be included, two along the light-cone direction 
and two along the transverse direction. The fully gauge invariant expression for $\hat{q}$ is now given as, 
\bea
\hat{q} &=& \frac{ 4 \pi^{2} \A_{s} }{N_{c}}  \int \frac{ d y^{-} d^{2} y_{\perp} d^{2} k_{\perp}}{(2\pi)^{3}} e^{i \frac{k_{\perp}^{2} y^{-}}{2 q^{-}} 
-i k_{\perp} \cdot y_{\perp} }  \nn \\
&\times& \left\langle P \left| \tr \left[ {F_{\perp}^{a}}^{+ \mu} (y^{-}, y_{\perp})
U^{\dag}(\infty^{-}, y_{\perp} ; 0^{-}, y_{\perp} ) \right. \right. \right. \nn \\ 
\ata T^{\dag}(\infty^{-},\vec{\infty}_{\perp};\infty^{-},y_{\perp})
T(\infty^{-},\infty_{\perp};\infty^{-},0_{\perp}) \nn \\ 
\ata \left. \left. \left. U(\infty^{-}, 0_{\perp} ; 0^{-}, 0_{\perp} ){ F_{\perp}^{b}}^{+}_{,\mu} \right] \right| P \right\rangle . \label{qhat_gaugeinvar}
\eea
In the equation above, $U$ represents a Wilson line along the $(-)$ light-cone direction and $T$ represents a
Wilson line along the transverse light-cone direction. If the calculation were being carried out in covariant 
gauge, only the light-cone Wilson lines will contribute, while for the calculation in light-cone gauge, only 
the transverse Wilson lines will contribute. Thus while the exact expressions are rather different in the 
two gauges, both may be derived from Eq.~\eqref{qhat_gaugeinvar}.
Given the extent of the Wilson lines (and the issues related with analytically continuing an euclidean operator product to one 
that is almost light-like separated), it appears almost 
impossible to evaluate these on a finite size lattice. 

However, there exists an alternative, based on the similarity between $\hat{q}$ and the gluon distribution function and 
the method by which parton distribution functions (PDFs) are evaluated on the lattice~\cite{Kronfeld:1984zv,Martinelli:1987zd,Martinelli:1988rr,Martinelli:1988xs}, 
i.e., using the method of operator product expansions. Imagine a high energy process e.g. the deep inelastic scattering (DIS) 
of an electron with momentum $k$ off a single quark prepared with momentum $p$, 
at one edge of a finite volume $V$ which is maintained at a fixed temperature $T\sim \Lambda_{QCD}$. At this temperature the 
volume will be filed with strongly interacting matter, which at temperatures somewhat below $\Lambda_{QCD}$ will be a hadronic 
gas and at very high temperatures will be quark gluon plasma. We maintain the chemical potential $\mu = 0$ so that the contents 
have the conserved charges of the vacuum. On scattering off the electron, the quark will produce a hard virtual quark 
which will then propagate through the medium. In vacuum such a parton would undergo a perturbative shower, spraying partons with ever lower 
virtuality until the scale becomes comparable to $\Lambda_{QCD}$ and hadronization begins to set in. In the presence of a strongly 
interacting medium the produced shower will scatter off the constituents in the medium, diffuse in transverse and longitudinal momentum, 
and be induced to radiate more partons leading to a further degradation in the energy of the part of the jet which escapes the medium.

If the medium is not larger than $E/\mu_{0}^{2}$, where $E$ is the energy of the jet, and $\mu_{0}$ is the minimum scale below 
which pQCD is no longer applicable, a portion of the jet will hadronize outside the medium. The differential cross section for any particular 
outcome from such a hard scattering process can
be expressed using the standard factorized formula,
\bea
d \sigma_{h} = \frac{ \A^{2} }{k \cdot p Q^{4}}\mathcal{L}_{\mu \nu} d W^{\mu \nu},
\eea
where $\mathcal{L}^{\mu \nu}$ is the usual leptonic tensor and $dW^{\mu \nu}$ is the differential 
hadronic tensor for the particular process of interest; all interactions which involve the QCD coupling $g$ are contained within the hadronic tensor.

Say further that in the hadronic tensor, we could factorize the initial distribution of the hard quark, the hard scattering off the photon and the 
final propagation through the medium as, 
\bea
d W^{\mu \nu}  = \int dx f(x) d \hat{\sigma}^{\mu \nu} D(\{p_{f}\}) .
\eea
In the equation above, $f(x)$ represents the distribution of the initial quark; in the case of quark inside a proton this would simply be the 
parton distribution function. In the case of a single quark it is simply $\kd(1-x)$. The term $\hat{\sigma}^{\mu \nu} $ represents the 
hard cross section for the scattering of a quark off a virtual photon. The function $D$ which is a function of the set of measured final state 
momenta $\{ p_{f} \}$ includes all final state effects after the hard collision of the quark with the photon. The general structure of $D$ may be 
written as
\bea
D( \{ p_{f} \}  )  &=& \sum_{j,k} \lc M |  \mathcal{O}_{j} | M \rc \nn \\
\ata \lc 0 | \mathcal{Q}^{\dag}_{j} | \{p_{f}\}  X \rc \lc \{p_{f}\} X | \mathcal{Q}_{j} | 0 \rc,
\eea
where, $| M \rc$ represents the medium where the jet interacts, $| \{p_{f}\} X \rc$ represents 
an inclusive hadronic state containing the detected hard momenta $\{p_{f}\}$ and other states which are not 
part of the medium. The operators $\mathcal{Q}, \mathcal{Q}^{\dag}$ represent the part of the 
process which occurs outside the medium and fragments to yield the detected ``non-medium'' final state. 
The remaining operator $\mathcal{O}_{j}$ represents the part of the process that occurs within the medium.
In a real heavy-ion collision such a distinction may be impossible to even formulate. However, in the 
theoretical scenario of a hard jet propagating through a finite medium, such a separation can be 
carried out order by order.

In the case of a single inclusive measured 
hadronic momentum, $D$ would become the standard fragmentation function (if $\mathcal{O}_{i} =1 $  there would be 
no medium effect, otherwise one would obtain the medium modified fragmentation function). 
For more exclusive observables (with more specified 
momenta), $D$ would represent a more complicated 
object~\cite{Majumder:2004wh}. 
We should make it clear that the momenta which specify $D$ do not need to be hadronic and may be completely partonic; in fact 
the particular $D$ that we will consider will be completely partonic.

In the remainder of this paper, we will consider evaluating $\mathcal{O}_{i}$ by perturbing in the weak coupling of the hard produced quark with
the medium. Note that this does not assume that the coupling within the medium is perturbatively weak. We will encode the effect 
of the medium on the hard quark in terms of an infinite series of local, power suppressed, operators (suppressed by powers of the hard scale $Q^{2}$).
Thus $\mathcal{O}_{i}$ will be obtained as a series of local operators $\mathcal{\overline{O}}^{i}_{n}$ and ever more suppressed perturbative coefficients $c^{i}_{n}/\left[Q^{2}\right]^{n}$,
\bea
\mathcal{O}_{i} = \sum_{n}  \frac{c^{i}_{n}}{\left[ Q^{2} \right]^{n}} \mathcal{\overline{O}}^{i}_{n}. \label{OPE}
\eea

While perturbation theory is valid for the interactions of the hard quark, it is not valid for  the local operator products. 
Any evaluation in perturbation theory necessarily requires the specification of a gauge and the calculations in this paper will 
be no different. Each choice of gauge will result in a slightly different set of perturbative terms along with a slightly different 
set of local operator products. For gauge invariant observables such as $D$ the total sum will be gauge invariant.
To demonstrate this however, one needs to be able to evaluate the operator products (for at least the first couple of 
terms). 

In all prior attempts to evaluate $D$, the non-perturbative sector has never been evaluated exactly. In the HT scheme, 
which is closest in spirit to the present discussion, the operator products (or some combination of them) are treated as 
parameters of the theory. A model is assumed for how they would depend on intrinsic properties of the medium such as 
the temperature $T$. The overall normalization is set by comparing with one data point. In this paper we present the 
first effort to estimate these operator products non-perturbatively on the lattice. 

The primary motivation for this effort is to test if such an approach is at all feasible. 
There is no attempt to be exhaustive and only the simplest process of jet broadening will be considered: the broadening 
of a single quark by a single scattering with momentum exchange $k_{\perp}$ in a hot medium. Dividing the mean $k_{\perp}^{2}$
by the length of the medium will yield the transport coefficient $\hat{q}$. The question that will be addressed in this paper is 
if such an approach is at all possible. To this end, we will calculate the perturbative part only in $A^{-} = 0$ gauge and the 
non-perturbative part in quenched $SU(2)$ lattice gauge theory. In this sense, this paper should be viewed as a ``proof of 
principle'' of such a methodology. Issues related to renormalization on both the perturbative and the non-perturbative 
side will be ignored. 
The evaluation of the perturbative coefficient functions in an alternate gauge, 
the computation of the modification of the shower pattern of the jet, and the 
evaluation of the non-perturbative operator products in $SU(3)$ will be left for future efforts.

We note in passing that, while in this paper, we assumed the factorization of the 
hard scatting from the final state scattering, this (assumption) is not strictly necessary in such a framework. Indeed one may 
consider \epem annihilation within such an enclosure and calculate the modification of the back-to-back pair of jets. 
Depending on the choice of observable and gauge this will lead to a unique expansion in the form of Eq.~\eqref{OPE}.

\section{Leading order derivation}

In this section, the operator expectation $D$ will be factorized into a perturbative and non-perturbative part. 
As pointed out above, we will consider the simplest process of jet broadening at leading order in the medium.  
To this end, we consider the propagation of a hard virtual quark through a hot medium with the quantum 
numbers of the vacuum. The large scale associated with this parton allow for the use of perturbation 
theory and we compute the first perturbative contribution which occurs only in the presence of a medium. 

Imagine a quark in a well defined momentum state $ | q \rangle \equiv |q^+, q^- ,0_\perp \rangle$ 
impinging on a medium $| M \rangle $ and then exiting 
in the state 
\[
| q + k \rangle \equiv \left| \frac{\left( k_\perp^2 + Q^2 \right)}{[2 (q^- + k^-) ] } , q^-+ k^- ,  \vec{k}_\perp \right\rangle,
\]
with the medium state absorbing this change in momentum and becoming $| X \rc$. The quark is assumed to 
be space-like off-shell with virtuality $Q^2 = 2 q^+q^- \leq 0$ with the negative $z$-axis defined as the direction of the 
propagating quark. In a physical situation, one would have a gluon radiated off a quark, with either the gluon or the 
quark space-like off-shell (or both). The space-like parton would be placed closer to its mass shell by scattering in the 
medium. The rate of scattering is controlled by the transport coefficient $\hat{q}$. To mimic this process we have considered
the very simple process of a space-like quark scattering off the glue field in an extended medium. The case of 
an on-shell quark is included in the limit of $Q^2 \ra 0$. 

Consider the reaction in the rest frame of the medium. In this frame $q^0>0$, and we have defined the $z$-axis such 
that $q_z < 0$. In this choice of frame, for a space-like quark we have $q^+ = (q^0 + q_z)/\sqrt{2} \leq 0$ and $q^- = (q^0 - q_z)/\sqrt{2} >0$. If the $z$-axis were chosen such that $q_z>0$ the $q^+$ and $q^-$ will simply switch roles. 
For a space-like quark we have $q^0 \leq | q_z |$, and this implies that $q^- > q^+$. For a jet one requires $q^{-} \gg q^{+}$. 
Alternatively stated $\sqrt{| q_{0}^{2} - q_{z}^{2} |}  \ll q_{0} \sim - q_{z}$.

\begin{figure}[htbp]
%\begin{center}
%  \epsfxsize 80mm
%\hspace{0cm}
\resizebox{2in}{1.5in}{\includegraphics{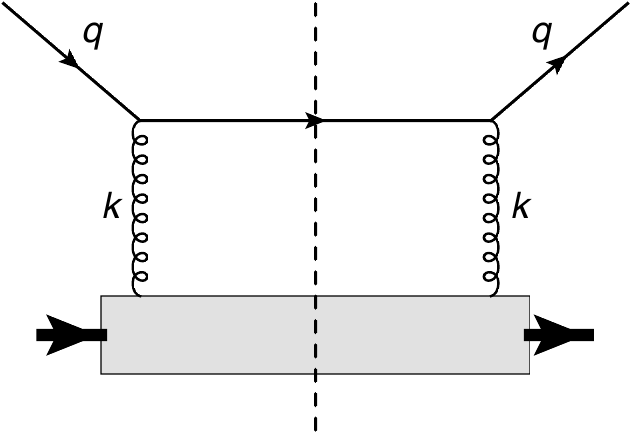}} 
%\vspace{0.25cm}
    \caption{ A quark scattering off a gluon in medium $|M \rc$.}
    \label{fig1}
%  \end{center}
\end{figure}

The spin color averaged transition probability (or matrix element) for this 
process, in the interaction picture,  is given as 
\bea
W(k) \!\!\!&=&\!\!\!  \frac{1}{2N_c } \langle q^- ; M |  T^* e^{ i \int_0^{t} d  t H_I(t)  } | q^-+k_\perp , X \rangle
\nn \\
\mbx \!\!\!\ata \!\!\! \langle   q^-+k_\perp , X  | T e^{-i \int_0^{t} dt H_I(t) }  | q^- , M \rangle ,
\eea  
where, we have averaged over the initial color and spin of the quark, assuming that the medium is in a fixed state. In the 
case of a thermal medium one may use the density matrix to average out the initial state. 
We will assume that all this is implicitly included in $|M\rc$. In the equation above, 
$H_I = \int d^3 x \psibar (x) i g t^a \g^\mu A^a_\mu (x) \psi(x)$ and $T$ ($T^*$) represents time(anti-time)-ordering. 
Expanding the exponential to leading order yields, 
\bea
W(k)  &=&  \frac{g^2}{2 N_c}\langle q^- ; M | \int d^4 x d^4 y \psibar(y) \f\! A(y) \psi(y) \nn \\ 
\ata | q^- + k_\perp ; X \rangle 
\langle q^- + k_\perp; X |  \nn \\
\ata \psibar(x) \f \!A(x) \psi(x) | q^- ; M \rangle, 
\eea
where, $A_\mu = t^a A^a_\mu$.  To deal with the factors of time $t$ and volume $V$, we introduce box normalization for the 
quark wave-functions and later take the limit of $t,V \ra \infty$.
In box normalization, $\psi(x) | q^- \rangle = e^{ - i q \cdot x} u(q)/\sqrt{V} $, we get,
\bea 
W(k) &=& 
\frac{g^2}{2 N_{c} V } \int d^4 x d^4 y \tr \left[ \langle M | \frac{\f q}{2 E_{q} } \f A(y) \right.   \label{DIS-like-propagator}\\
\ata \left.  {\rm Disc} \left[ \frac{(\f q + \f k)}{ (q+k)^{2}  + i \e}  \right] \f A(x) | M \rangle \right] e^{-ik \cdot (y - x)} . \nn 
\eea
Shifting, the $x$ and $y$ integrations, the four volume may be extracted ($\int d^{4} x = t V$) 
and divided out by the factors in the denominator. The mean $k^2_\perp$ 
which yields $\hat{q}$ has the obvious definition, 
\bea
\hat{q} = \sum_{k}  k_\perp^2  \frac{W(k)}{t},
\eea
where, we have summed over all values of the four vector $k$ with the restriction that 
the final out going quark remain on shell. Where $t$ represents the time spent by the 
hard quark in the thermal volume $V$. 
With the overall factor of four-volume removed we can take $t,V \ra \infty$.

We will now demonstrate that in the limit that $q$ goes near on-shell, i.e., $q^{-} \gg q^{+}$, the expression 
above reduces to the well known expression for the transport coefficient $\hat{q}$. Taking the limit that 
$Q^{2} = 2 q^{+} q^{-} \ra 0$ while $q^{-} \ra \infty$, we can simplify the Dirac trace as
\bea
&& \lc M | \tr[\f q  \f A (\f q + \f k) \f A] |M \rc \nn \\
&=& 8 ( q^- )^2 \tr[t^a t^b] \lc M | A_a^+ (y) A_b^+ (x) | M \rc.
\eea
The imaginary part of the propagator yields the on-shell $\kd$-function, which may also be simplified as, 
\bea
\kd [ (q+k)^2] \simeq \frac{1}{2q^-} \kd \left( k^+ - \frac{k_\perp^2}{2q^-} \right). \label{delta}
\eea
Since, $k^-$ has been ignored, compared to $q^-$ it may be integrated over to yield $2 \pi \kd( y^+)$.
The $k_\perp^2$ may be combined 
with the vector potentials to yield, $\nabla_\perp A^+ \simeq F_\perp^+$. Absorbing both factors of $k_\perp$, 
we obtain an expression containing only field strength tensors.

Substituting the above simplifications, one obtains, 
\bea
\hat{q} &=& \frac{4 \pi^2 \A_s}{ N_c } \int \frac{dy^- d^{2} y_{\perp}}{(2 \pi)^{3}} d^{2} k_{\perp} 
e^{ -i \frac{k_{\perp}^{2}}{2q^{-} } \x y^{-} +  i\vec{k}_{\perp} \x \vec{y_{\perp} } } \nn \\
\ata \lc M | {F^{+,}}_\perp (y^-,y_{\perp}) F_\perp^+ (0)  | M \rc.
\eea
This is the standard definition of $\hat{q}$. Note that nothing is specified about $|M\rc$, it may indeed be an 
arbitrary medium. If $|M \rc$ is a thermal medium, then it must be averaged over 
in the sum over all initial states. Averaging with a Boltzmann weight will yield, 
\bea
\hat{q} = \frac{4 \pi^2 \A_s}{ N_c } \int \frac{dy^- d^{2} y_{\perp}}{(2 \pi)^{3}} d^{2} k_{\perp} 
e^{ -i \frac{k_{\perp}^{2}}{2q^{-} } \x y^{-} +  i\vec{k}_{\perp} \x \vec{y_{\perp} } } \nn \\
\lc n |\frac{e^{-\B E_n}}{Z} {F^{+,}}_\perp (y^-) F_\perp^+ (0)  | n \rc.
\eea

Note that in the above derivation, no ordering is introduced between the two field strength operators. 
The expression above is not gauge invariant, but is gauge covariant. This implies, that if one were to carry out an 
operator product expansion in terms of local operators, one could reorganize the expansion to only contain 
gauge invariant local operators. Any gauge dependence would then only be contained in the coefficient functions.  

\section{The off-shell regime and the non-physical regime.}

In the preceding section, we considered the process of a near on-shell quark propagating through a 
hot medium, at leading order in the scattering off the medium. In this section, the case of a 
slightly off-shell quark will be considered. The quark virtuality or offshellness will still be small 
compared to the energy. Once the operator products have been isolated, we will consider the 
process in the region of very high virtuality, of the order of the energy, and consider an expansion 
in a power series with increasing negative powers of the virtuality.

Consider the imaginary part of the propagator in Eq.~\eqref{DIS-like-propagator}. In the limit that $q^{-}$ is very large, 
and $q^{+}$ is vanishingly small, 
there is a pole at the point where $k^{+} = (k_{\perp}^{2} )/(2 q^{-})$. 
In the regime where $q^{+} \ll q^{-} $ but $q^{+}$ is not vanishingly small (i.e., the parton has a non-negligible virtuality) 
we will obtain small additive contributions to the gauge 
covariant structure derived above. In this section we consider the more physical limit where 
$q^{+} q^{-} \sim k_{\perp}^{2} \sim \lambda^{2} (q^{-})^{2}$, where $\lambda$ is a small dimensionless constant.
In this case, the Dirac matrix structure will be simplified by taking the trace as, 
\be
\tr[\f q \f A(0) (\f q + \f k) \f A(y) ] 
= 4 A^{\mu} (0) G_{\mu \nu} A^{\nu}(y),  \nn 
\ee \\
with, $G_{\mu \nu} = \left[ q^{\mu} (q+k)^{\nu} + q^{\nu} (q+k)^{\mu} - (q+k) \x q g_{\mu \nu}   \right]$
Expanding this out, we obtain, 
\bea
&& A(0)\x G \x A(y) \nn \\
&=& 2 q^{-} A^{+}(0) q^{-} A^{+}(y) + q^{-} A^{+}(0) (q^{+} + k^{+}) A^{-}(y)  \nn \\
&+&  q^{+} A^{-}(0) q^{-} A^{+} + (q^{+} + k^{+}) A^{-}(0) q^{-} A^{+}(y)  \nn \\
&+&   q^{-}A^{+}(0) q^{+}A^{-}(y) +   2q^{+} (q^{+} + k^{+} ) A^{-}(0) A^{-}(y) \nn \\
&-& q^{-}A^{+}(0) k_{\perp}\x A_{\perp}(y) - k_{\perp}\x A_{\perp}(0) q^{-} A^{+}(y) \nn \\
&-& q^{+}A^{-}(0) k_{\perp}\x A_{\perp}(y) - k_{\perp}\x A_{\perp}(0) q^{+} A^{-}(y) \nn \\
&-& [q^{-} (q^{+} + k^{+} ) + q^{+}q^{-} ]  \\
\ata[ A^{+}(0) A^{-} (y) + A^{-}(0) A^{+}(y) - A_{\perp}(0) \x A_{\perp} (y) ].   \nn 
\eea

We now consider this expression in $A^{-}=0$ gauge, where we may drop terms which go as $Q^{2}/q^{-} \sim \lambda^{2} q^{-}$. This leads to a considerable 
simplification of the final expression,  
\bea
&& A\x G \x A = 2 q^{-} A^{+}(0) q^{-} A^{+}(y)   \nn \\  
&+& q^{-}A^{+}(0) k_{\perp, \mu } \x A_{\perp}^{\mu}(y) + k_{\perp, \mu}\x A_{\perp}^{\mu}(0) q^{-} A^{+}(y) \nn \\
&-& [q^{-} (k^{+}+q^{+})  + q^{-} q^{+} ] [  A_{\perp, \mu}(0) \x A_{\perp}^{\mu}(y) ] . \label{light-cone-terms}
\eea
The exponential phase factor is, 
\bea
e^{i \phi} = \exp \left[ i \left\{ \left( \frac{k_{\perp}^{2}}{2q^{-} } - q^{+} \right) y^{-}   +  k_{\perp, \mu} y_{\perp}^{\mu} \right\} \right] ,
\eea
where the general ($\perp$)-4-vector implies $A_{\perp} \equiv [0,0,\vec{A}_{\perp}]$. Using these relations, we may 
simplify,
\bea
&& 2 (q^{-})^{2} ( - k_{\perp}^{\mu} k_{\perp, \mu} ) A^{+}(0) A^{+}(y)e^{i\phi(y)} \label{term-1}\\ 
&=& - 2(q^{-})^{2} \nabla_{\perp}^{\mu} A^{+} (0) \nabla_{\perp, \mu} A^{+}(y) e^{i \phi(y)}. \nn
\eea
The next set of terms simplify as, 
\bea
&& e^{i\phi} q^{-} A^{+}(0) k_{\perp , \mu} A_{\perp}^{\mu} (y) k_{\perp}^{2} \label{term-2} \\ 
&=& 2 (q^{-})^{2} i \nabla_{\perp , \mu} A^{+}  \left[ q^{+}  - i \prt^{+} \right] A_{\perp}^{\mu} (x) \nn \\
&=& 2 (q^{-})^{2} \left[  \nabla_{\perp , \mu} A^{+}  \prt^{+}  A_{\perp}^{\mu} (x) + i  \nabla_{\perp , \mu} A^{+}q^{+}  A_{\perp}^{\mu} (y) \right].  \nn
\eea
The first term in the bracket above, can be combined with Eq.~\eqref{term-1} to produce the field strength tensor at location $x$.
There is another term similar to the one above which can be combined to form the field strength tensor at the origin.
The last line in Eq.~\eqref{light-cone-terms} may be re-expressed as, 
\bea
&&\!\!\!\!\!\! - 2 (q^{-})^{2}  \left[ \prt^{+} A_{\perp,\mu}(0) \prt^{+} A_{\perp}^{\mu} (y) + 2 i q^{+} A_{\perp, \mu} \prt^{+} A_{\perp}^{\mu}(y)  \right. \nn \\
&& -\left. iq^{+} \prt^{+} A_{\perp, \mu} (0) A_{\perp}^{\mu}  + 2 (q^{+})^{2} A_{\perp,\mu}(0) A_{\perp}^{\mu} (y) \right]. \label{term-3}
\eea 

The first set of terms in the equations above [Eqs.(\ref{term-1},\ref{term-2},\ref{term-3})] can be combined to obtain the known form that appears 
in the definition of the on-shell $\hat{q}$, i.e. $2(q^{-})^{2} F^{+}_{\perp, \mu} F^{\mu, +}_{\perp} $. 
Note that all terms in Eq.~\eqref{term-3} are rather small [they scale as $\lambda^{2} 2 (q^{-})^{2} \nabla_{\perp,\mu}
A^{+}(0) \nabla_{\perp}^{\mu} A^{+}(y)$] and 
thus the remaining terms may be ignored.

We now have an expression for the transport coefficient $\hat{q}$ over a range of values of $q^{+}$ where $q^{+} \ll q^{-}$ but is still 
large enough that $Q^{2} = 2q^{+} q^{-} \gg \Lambda_{QCD}^{2}$. We can now take this particular operator product and 
consider its behavior over the entire complex plane of $q^{+}$.

We now analytically continue to the region where 
$q^{+ } < 0 $ and $|q^{+}| \sim q^{-} \gg k$. Consider the analytically continued, unphysical expression, 
\bea
\hat{Q} &=& \frac{4 \pi^{2} \A_{s}}{N_{c}}\!\!\! \int \frac{d^{4}y d^{4} k}{(2\pi)^{4}} 
e^{i k \x y} \frac{2 (q^{-})^{2}}{ \sqrt{2} q^{-} } 
\nn \\
\ata \frac{ \lc M | F^{+ \perp}(0) 
F_{\perp,}^{+}(y) | M \rc}{ (q+k)^{2} + i \e } .
\eea
We introduce a new object $\hat{Q}$ to indicate that the expression above is not the jet 
transport coefficient $\hat{q}$. The discontinuity of the above expression in the region 
$-q^{- } \ll q^{+} \ll q^{-}$ corresponds to $\hat{q}$.

In the regime where $q^{+} \sim q^{-} \gg k$, one can expand out the denominator as, 
\bea
\frac{1}{( Q^{2}  - k_{\perp}^{2}  + 2 q \x k  )} 
\simeq \frac{1}{Q^{2}} \sum_{n=0}^{\infty}\left(  \frac{ - 2q \x k  + k_{\perp}^{2} }{ Q^{2} }   \right)^{n}\!\!\!.
\eea
The instances of the gluon momentum $k$ may be replaced with derivatives. 
Adding, gluon scattering terms, we can convert the regular derivatives into covariant derivatives. 
Thus we obtain a  series of gauge covariant expressions for the jet transport  coefficient. 
\bea
\hat{Q} &=&  \frac{4 \pi^{2} \A_{s}}{N_{c}}\!\!\! \int \frac{d^{4}y d^{4} k}{(2\pi)^{4}} e^{i k \x y}  \frac{\sqrt{2}q^{-}}{Q^{2}}  \\
\ata \!\!\!\lc M | F^{+ \mu}_{\perp}(0) \sum_{n=0}^{\infty} \left( \frac{ -q\x i\mathcal{D} - \mathcal{D}_{\perp}^{2} }{Q^{2}} \right)^{n} F^{+}_{\perp, \mu}  (y) | M \rc . \nn
\eea
With all instances of $k$ removed from the integrand (except for the phase factor), the integrals over all components of
 $k$ can be carried out to yield four $\kd$-functions over the position $y$. This yields a very simple expression 
 for $\hat{q}$ in $A^{-} = 0$ gauge, in terms of local gauge invariant operators, 
 \bea
 \hat{Q} &=& \frac{4\sqrt{2} \pi^{2} \A_{s} q^{-}}{N_{c}  Q^{2}} \nn \\
 &\times & \lc M | F^{+ \mu}_{\perp}
 \sum_{n=0}^{\infty} \!\!\left(\!\! \frac{ -q\x i\mathcal{D} - \mathcal{D}_{\perp}^{2} }{Q^{2}} \!\!\right)^{n} 
 \!\!\!F^{+}_{\perp, \mu}  | M \rc. \label{Qhat-OPE}
 \eea

The above expression requires some discussion. The discontinuity in the expression above 
across the real axis of $q^{+}$ corresponds to the transport coefficient $\hat{q}$ when $-q^{-} \ll q^{+} \ll q^{-}$. 
For $q^{+} \sim q^{-}$ and positive, there is another source of a discontinuity, from real hard gluon 
emission. This part is perturbatively calculable as long as $Q^{2} = 2 q^{+} q^{-} \gg \Lambda_{QCD}^{2}$ 
and does not depend on any properties of the medium. In the region where $Q^{2}$ is space-like or 
$q^{+} \ll  -\Lambda_{QCD}$ there is no discontinuity across the real axis. 
Alternatively speaking, in the deep space-like region 
the internal quark-line cannot go on-shell. For virtualities which are not in the deep space-like region, 
the quark can still absorb a gluon from the medium and go on-shell and there will be a discontinuity.

\section{Dispersion relations}

In the preceding section, the expression for $\hat{q}$ was generalized to the region of (a physically realizable) non-zero virtuality and 
then considered in the region of (unphysical) very high virtuality. In the current section the two expressions will be 
related via dispersion relations in the complex $q^{+}$ plain. The expansion in the unphysical region [Eq.~\eqref{Qhat-OPE}]  will 
be used to estimate the value of $\hat{q}$ in the physical region.  

In order to evaluate $\hat{q} = \dsc \left[  \hat{Q} \right]$ for $q^{+} \sim \lambda^{2} q^{-}$, 
we will use the method of dispersion relations: We will evaluate a similar integral in a region of the $q^{+}$ complex plain where 
there is no discontinuity and use methods of contour integration to relate the evaluated 
integral to $\hat{q}$.

Consider the integral,
\bea
I_{m} = \oint \frac{d q^{+}}{2 \pi i} \frac{  \hat{Q}(q^{+}) }{ \left( q^{+}  + Q_{0} \right)^{m} },
\eea
where $Q_{0}$ is large and positive. The contour is taken as a small counter-clockwise circle 
around the point $q^{+} = - Q_{0}$. The residue of this integral is given as, 
\bea
I_{m} = \left. \frac{ d^{m-1} }{ d^{m-1} q^{+}  } \hat{Q}(q^{+}) \right|_{q^{+} = - Q_{0}} .
\eea

While this analysis can be carried out for arbitrary $m$, we consider, for definiteness, the 
case of $m=1$. 
In the limit where $ |q^{+} |\gg \lambda Q$, we obtain Eq.~\eqref{Qhat-OPE} with $Q^{2}$  replaced by 
$- 2 q^{-} Q_{0}$, i.e.
\bea
I_{1} \!\!= \! \frac{4\sqrt{2} \pi^{2} \A_{s} \lc M | F^{+ \mu}_{\perp} \!
 \sum\limits_{n=0}^{\infty} \!\!\left(\!\! \frac{ -q\x i\mathcal{D} - \mathcal{D}_{\perp}^{2} }{2 q^{-}Q_{0}} \!\right)^{n} 
 \!\!\!F^{+}_{\perp, \mu}  | M \rc }{N_{c} 2 Q_{0}}   . 
\eea
Since $q^{+},q^{-} \gg k_{\perp}^{2}$, the above operator relation in simplified as, 
\bea
I_{1} &=& \frac{ 4 \sqrt{2} \pi^{2} \A_{s}}{N_{c} 2 Q_{0}} \lc M | F^{+ \mu}_{\perp}
 \sum_{n=0}^{\infty} \!\!\left(\!\! \frac{ -i\mathcal{D}^{+} }{2 Q^{0}} + \frac{-i\mathcal{D}^{-}  }{2 q^{-}} \!\!\right)^{n} \nn \\
& \times & F^{+}_{\perp, \mu}  | M \rc.  \nn \\
&=&  \frac{ 4 \sqrt{2} \pi^{2} \A_{s}}{N_{c} 2  Q_{0}} \lc M | F^{+ \mu}_{\perp}
 \sum_{n=0}^{\infty}\sum_{m=0}^{n} {n \choose m} \left( \frac{ -i\mathcal{D}^{+} }{2 Q^{0}} \right)^{m} \nn \\ 
&\times& \left( \frac{-i\mathcal{D}^{-}  }{2 q^{-}} \right)^{n-m}   
F^{+}_{\perp, \mu}  | M \rc.  \nn  \\
&=& \frac{ 4\sqrt{2} \pi^{2} \A_{s}}{N_{c} 2 Q_{0}} \lc M | F^{+ \mu}_{\perp} \sum_{m=0}^{\infty} 
\frac{1}{m!}\left( \frac{-i \mathcal{D}^{+}}{ 2 Q^{0}} \right)^{m} \nn \\
&\times & \sum_{k=0}^{\infty} \frac{(m+k)!}{k!} \left(\frac{ - i \mathcal{D}^{-} }{2 q^{-} } \right)^{k}   F^{+}_{\perp, \mu}  | M \rc. \label{I1-expanded}
\eea

We can now deform the contour and evaluate it over the branch cut from $q^{+} > - \lambda^{2} Q $ to $q^{+} \ra \infty$.
This yields,
\bea
I_{1} &=& \frac{4 \pi^{2} \A_{S}}{N_{c}}  \int d q^{+} \frac{d^{4} y d^{4} k}{ (2\pi)^{4} } e^{ik \x y} 
\frac{\kd \left( k^{+} + q^{+} - \frac{k_{\perp}^{2}}{ 2 q^{-}} \right) }{ 2 q^{-} } \nn \\
\ata  \frac{ \lc M | F^{+ \mu} (0) F^{+}_{\mu , }(y)  | M \rc }{ \left( q^{+} + Q_{0} \right)  } \nn \\
&=& \int_{-\lambda^{2} Q}^{\lambda^{2} Q} d q^{+}  \frac{\hat{q} (q^{+})}{ q^{+} + Q_{0}} + \int_{0}^{\infty} d q^{+} V(q^{+}). 
\eea
The second term in the equation above, refers to the contribution to the operator above from vacuum gluon radiation, 
i.e., the Bremsstrahlung radiation of gluons from an off-shell quark. As such, it contributes only in the region where 
the virtuality of the incoming quark is time-like and is independent of the temperature of the medium. Thus for a 
fixed $T$ the second term above is a constant, while the first depends on the temperature of the 
medium.

The limits on the first integral in the equation above allow for a simple expansion of the denominator. The factor 
$Q^{0} \sim Q$ is much larger than the $q^{+} \sim \lambda^{2} Q$ in this region and thus we obtain, the much 
simplified relation, 
\bea
\int dq^{+} \frac{\hat{q}(q^{+})}{Q_{0}}  \sum_{n=0}^{\infty} \left[ \frac{-q^{+}}{Q_{0}} \right]^{n}  \simeq  I_{1}\! - \!\int_{0}^{\infty}\!\!\!\! d q^{+} V(q^{+}).
\eea
To obtain $\hat{q}$, a general functional form in the vicinity of $-\lambda^{2}Q \leq q^{+} \leq \lambda^{2} Q $ must be used. 
We start with the assumption that $\hat{q}$ at a fixed $q^{-}$ is a slowly varying function of $q^{+}$. This allows us to use a 
truncated Taylor expansion for $\hat{q}$ [We should point out that using the first few terms of the Taylor expansion is, in itself, an 
assumption regarding the functional form of $\hat{q} (q^{+})$]. To  provide a simple illustration of the procedure, we take only 3 terms; in the 
final numerical results we will only use those results where the first term greatly dominates over all 
subsequent terms (Note that an arbitrary number of terms in the Taylor expansion may be 
retained for a more accurate determination of $\hat{q}$), 
\bea 
\hat{q} (q^{+}) = \hat{q} + \hat{q}' q^{+} + \frac{\hat{q}'' (q^{+})^{2}}{2}. 
\eea
In the above equation $\hat{q}' = \prt \hat{q} / \prt q^{+} |_{q^{+} = 0}$.

Using the above truncated Taylor expansion we obtain, 
\bea
I_{1}   &=&   \frac{ \!\!\!\! \int\limits_{-Q^{+}}^{\,\,\,\,\,\,Q^{+}} \!\!\!\!\!dq^{+}  
\left[ \hat{q} + \hat{q} \left(\frac{q^{+}}{Q^{0}}\right)^{2} - \hat{q}' \frac{(q^{+})^{2}}{Q^{0}}  +  \hat{q}''  \frac{(q^{+})^{2} }{2} \right] }{Q^{0}}
\nn \\
&+& \int_{0}^{\infty} \!\!\!\!dq^{+} V(q^{+}) = \frac{2 \hat{q} Q^{+} }{Q_{0}} + \frac{ \hat{q}'' (Q^{+})^{3} }{3 Q_{0}}   \nn \\ 
& - &\frac{\hat{q}' 2 (Q^{+})^{3} }{ 3 Q_{0}^{2}} 
+ \hat{q} \frac{2 (Q^{+})^{3}}{3 Q_{0}^{3} } + \hat{q}'' \frac{ (Q^{+})^{5} }{ 5 Q_{0}^{3} } .   \label{I1-Taylor}
\eea
In the equation above, $Q^{+}$ represents the limit of integration over $q^{+}$ for the jet. For 
a jet with maximum virtuality $\mu^{2}$ and $(-)$ momentum $q^{-}$, $Q^{+} = \mu^{2}/(2q^{-})$.
One may now simply compare with the expression for $I_{1}$ from Eq.~\eqref{I1-expanded} and equate the vacuum subtracted coefficients of $Q_{0}^{n}$.

The methodology outlined above can be made even more precise and straightforward by setting a definite value for 
$Q^{0} = q^{-}$. While this will readjust the relative importance of the various terms in the series it allows for simpler 
set of operators that need to be evaluated numerically.  This simplifies $I_{1}$ in Eq.~\eqref{I1-expanded} to, 
\bea
\mbx \!\!\!I_{1}\!\! &=& \!\! \frac{ 2 \sqrt{2} \pi^{2} \A_{s}}{N_{c}  q^{-} } \lc M | F^{+ \mu}_{\perp}
 \sum_{n=0}^{\infty} \!\!\left(\!\! \frac{ -i\mathcal{D}^{0} }{q^{-}}  \!\!\right)^{n}  F^{+}_{\perp, \mu}  | M \rc,  \label{I1-simple}
\eea
and similarly simplifies Eq.~\eqref{I1-Taylor} with $Q_{0}$ replaced by $q^{-}$. For a virtuality $\mu^{2}$ such that 
$\Lambda_{QCD}^{2} \ll \mu^{2} \ll (q^{-})^{2}$, we can define a $q^{+}$ or virtuality averaged $\hat{q}$ as, 
\bea
\hat{\bar{q}} (Q^{+}) 2Q^{+}  &=& \int_{-Q^{+}}^{Q^{+}} dq^{+} \hat{q}(q^{+}) \nn \\
&\simeq & 2 \hat{q} Q^{+} + \frac{ \hat{q}'' (Q^{+})^{3} }{3},
\eea 
where the second line is only valid in the limit that $\hat{q}$ is a slow function of $q^{+}$
(or alternatively stated $Q^{+} \ll q^{-} $). We can obtain an estimate of this by studying the 2nd term in Eq.~\eqref{I1-simple}.
If this term is comparable to the first term then the above approximation is no longer valid. 
If this term is small, then one may obtain a good estimate of $\hat{q}$ from just the first term in the 
series in Eq.~\eqref{I1-simple}.
In the subsequent section the forms of the operators and their evaluation on the lattice will be discussed. 

\section{Lattice calculations}

In the preceding sections, the jet transport parameter $\hat{q}$, as obtained in the physical regime of jet momenta $q^{+} \sim \lambda^{2} q^{-} \ll q^{-}$, 
was related via dispersion relations to a series of local operators in an unphysical regime where $q^{+} = -q^{-}$. The 
availability of a series of local operators, suppressed by powers of the hard scale $q^{-}$ allow for the calculation of 
such non-perturbative operator products on the lattice. In essence, our task is to compute the finite temperature Minkowski space correlator, 
\bea
\mathcal{D}^{>} (t) = \sum_{n} \lc n | e^{-\beta H} \mathcal{O}_{1}(t) \mathcal{O}_{2} (0) | n \rc, 
\eea
in the limit that $t \ra 0$. In the equation above, $\beta$ is the inverse temperature ($\beta = 1/T$),  $H$ is the 
Hamiltonian operator, and $|n\rc$ represents an eigenstate of the Hamiltonian. 
Using the standard relations of the imaginary time formalism of finite temperature field theory, we can 
relate the Minkowski correlator with the Matsubara correlator in Euclidean space, 
\bea
\mathcal{D}^{>} \!( -i \tau) = \Delta(\tau)  =  \tr \!\left[  e^{-\!\int \limits_{0}^{\beta} \!d\tau  H(\tau) }  \! \mathcal{O}_{i}(\tau)  \mathcal{O}_{2}(0) \right] \!\! ,
\eea
for the case where there are no time derivatives in $\mathcal{O}_{1}$ and $\mathcal{O}_{2}$ and yields,
$
\mathcal{D}^{>} \!( -i \tau) = i^{N_{t}} \Delta(\tau)  
$
for a total of $N_{t}$ time derivatives in $D^{>} (t)$. As a result, we obtain the simple relation that 
\bea
\mathcal{D}^{>} (t=0) = i^{N_{t}} \Delta(\tau=0).
\eea
Using the above relation, the local operator products in Minkowski space may be obtained from the local operators in 
Euclidean space.

In the following, we list out the operators that must be evaluated and re-express them in a form where 
they may be easily calculated on the lattice. In this first exploratory attempt, the calculation will be carried out for an 
$SU(2)$ gauge theory on a space-temperature lattice in the simplified quenched approximation. Quark-less $SU(2)$ 
possesses a negative $\beta$-function as in full $QCD$. Since issues of higher order contributions and renormalization 
were ignored in the perturbative sector, renormalization will be dealt with in a very simplified fashion, in the non-perturbative sector. 
The extension to more sophisticated simulations in quenched (or unquenched)
$SU(3)$ will be left for future efforts. In defense of the current effort, we point out that in the context of jet transport coefficients in heavy-ion collisions, 
quenched calculations may provide a very realistic estimate, 
as the early dense plasma is believed to be gluon dominated.

In the language of links, the field strength tensor $t^{a}F^{a}_{\mu \nu}$ may be 
expressed as, 
\bea
F_{\mu \nu} \equiv t^{a} F^{a}_{\mu \nu} &=& \frac{ U_{\mu \nu}  -  U^{\dag}_{\mu \nu}  }{ 2 i g a_{L}^{2}},
\eea
Where, $U_{\mu \nu}$ represents a plaquette in the $\mu \nu$ plane and $a_{L}$ is the lattice spacing. 
Similarly, terms with a covariant derivative may be expressed as, 
\bea
\mbx\!\!\!\!\!
\mathcal{D}_{4} F_{\mu \nu} (x) \!=\!  \frac{ F_{\mu \nu} (x^{4} \!\!+\!\! a_{L},\vec{x}) - U_{4} (x^{4}\!\!,\vec{x} ) F_{\mu \nu} ( x^{4}\!\! ,\vec{x} ) }{ a_{L}  } ,
\eea
where, $U_{4}$ represents a gauge link in the $4$-direction. 
In this paper, we have only used the right derivative as we seek only an order of magnitude estimate of terms 
with a time derivative, as argued below.

The first operator to be evaluated is 
\bea
\lc M | F_{\perp}^{+ \mu}  {F_{\perp}}^{+}_{\mu} | M \rc &=& \sum \lc n | e^{- \beta H} F_{\perp}^{+  \mu} {F_\perp}^{,+}_{\mu} | n \rc  \\
&\equiv& \sum e^{-\beta E_{n}}  \lc n |  F_{\perp}^{+ \mu}   {F_{\perp}}^{,+}_{\mu} | n \rc, \nn
\eea
where, $|n\rc$ represents an eigenstate of the full Hamiltonian. 
We do not indicate the location of the two $F$ field strength tensor insertions as both are at the same location. 

We now discuss the rotation of the operator products to Euclidean space. This involves the two rotations:
\bea
&& x^{0} \ra -i x^{4}  \,\,\,{ \rm and } \,\,\,  A^{0} \ra i A^{4} \nn \\
&& \Rightarrow F^{0 i} \ra i F^{4 i}. 
\eea
As a result,
\bea
\mbx \!\!\! \lc [F^{0 1} \!\!+\! F^{3 1} ] [F^{01} \!\!+\! F^{31} ] \rc \ra \lc F^{31}\! F^{31} \rc  -  \lc F^{41}\!F^{41} \rc. \label{no-cross}
\eea
In the above equation, we have ignored terms such as ($F^{31} F^{41}$), as their vacuum subtracted contributions 
turn out to be rather small in the region where we will attempt to estimate $\hat{q}$ [this is plotted in Fig.~\eqref{fig3} and will be discussed below].

In the following, we will first discuss the lattice calculation of the operator $\sum_{i=1,2}( F^{3i} F^{3i} - F^{4i}F^{4i} )$.
The reader will note that, with the addition of the extra term $( F^{21}\! F^{21}  -  F^{43}\!F^{43} )$, this will become the 
operator for the entropy density (up to normalization constants). For an isotropic lattice, one could even estimate the value of  $\sum_{i=1,2}( F^{3i} F^{3i} - F^{4i}F^{4i} )$ 
as $2/3$ times the 
entropy density. In our calculation, the jet travels in the $z$-direction. As a result, the entire problem (perturbative and non-perturbative sectors) 
is not isotropic, even though the lattice part of the calculation is isotropic. Also the remainder of the operators required for the calculation of $\hat{q}$ 
have no simple relation with well known operators. Hence, we will directly evaluate the operator mentioned above and not try to estimate its 
value from the known results of the entropy density. 
The notion that $\hat{q}$ may be proportional to the entropy density has been prevalent in jet modification phenomenology and has been used in 
various calculations of jet modification, e.g., see  Ref.~\cite{Majumder:2011uk}.

We report results on a $(4\times n_{t})^{3}\times n_{t}$ lattice where $n_{t}$ is varied from $3$ to $6$. Note that finite temperature 
calculations are meant to be carried out in the limit that $n_{t} \ll n_{s}$. 
For $n_{t} = 2$ we have also carried out a calculation with $n_{s} = 12$ and $n_{s}=8$, these have not been presented as they 
show very little variation with $n_{s}$ for a fixed $n_{t}$.  
We have not repeated the calculation with smaller values of $n_{t} $ and the largest value of $n_{s}=24$ as the  
results for $n_{t} =2,3$ do not seem to be have any dependence on $n_{s}$ for $n_{s}> 12$.

\begin{figure}[htb!]
%\begin{center}
%  \epsfxsize 80mm
%\hspace{0cm}
\resizebox{3in}{3in}{\includegraphics{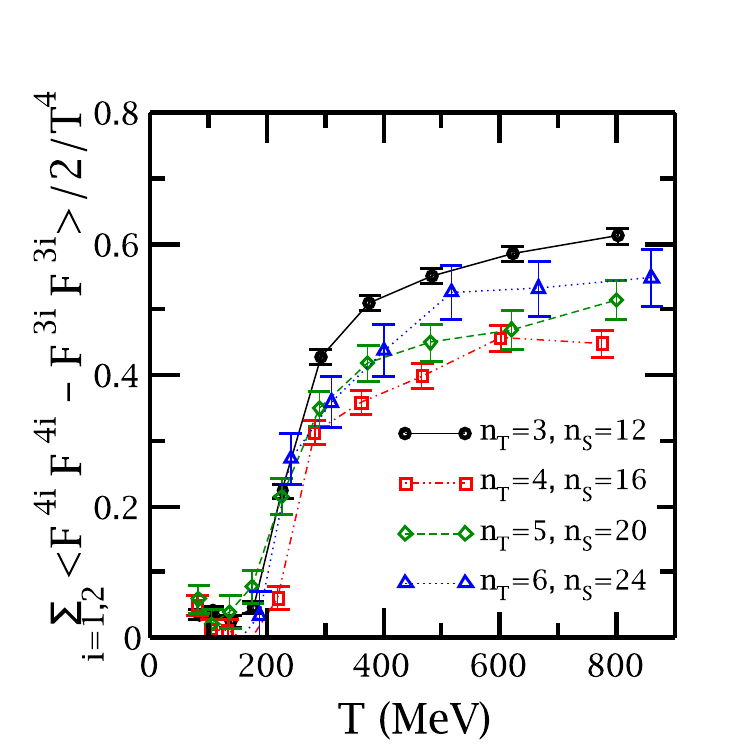}} 
%\vspace{0.25cm}
    \caption{(Color online) The temperature dependence of the local operator $\lc F^{+ i} F^{+ i} \rc$, scaled by $T^{4}$ to make it 
    dimensionless. The lattice spacing is set using Eq.~\eqref{lattice-spacing}. The expectation of the operator product shows a 
    transition in the vicinity of $T \sim 250-350$~MeV. See text for 
    details.}
    \label{fig2}
%  \end{center}
\end{figure}

For this first attempt we will use the Wilson gauge action for $SU(2)$~\cite{Engels:1980ty,Creutz:1984mg}. 
The scale (or lattice spacing) is set on the lattice using two different renormalization 
group formulas: The first is based on the two loop perturbative RG equation for the 
string tension~\cite{Engels:1980ty,Creutz:1984mg}, which yields the following formula for the lattice spacing,
\bea
a_{L} = \frac{1}{\Lambda_{L}} \left(\frac{11 g^{2}}{ 24 \pi^{2}}\right)^{-\frac{51}{121}} \exp \left( - \frac{12\pi^{2} }{11 g^{2}} \right), \label{lattice-spacing}
\eea
where, $g$ represents the bare lattice coupling and $\Lambda_{L}$ represents the one 
dimension-full parameter on the lattice. Comparing with the vacuum string tension, we have used $\Lambda_{L}=5.3$~MeV.   
For a lattice at finite temperature or one with $n_{t} \ll n_{s}$, the temperature is obtained as
\bea
T = \frac{1}{n_{t} a_{L}}.
\eea
The results for the field-strength-field-strength correlation $\sum_{i=1,2}( F^{3i} F^{3i} - F^{4i}F^{4i} )/2$  with this choice of 
formula for the lattice spacing are presented in Fig~\ref{fig2}. The resulting correlation is scaled by $T^{4}$ as obtained 
from the formula above.

\begin{figure}[htb!]
%\begin{center}
%  \epsfxsize 80mm
%\hspace{0cm}
\resizebox{3in}{3in}{\includegraphics{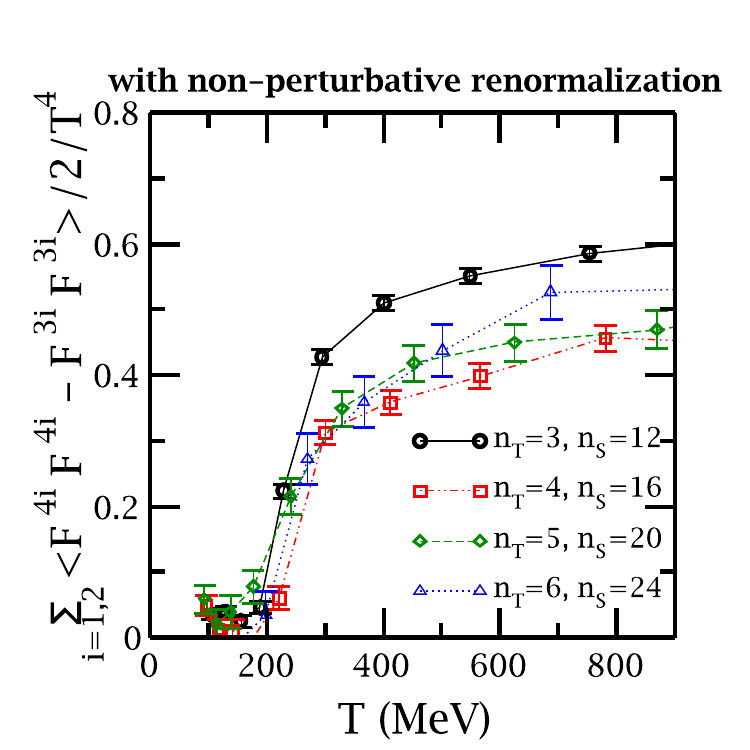}} 
%\vspace{0.25cm}
    \caption{(Color online) Same as Fig.~\ref{fig2}, except for the use of non-perturbative RG factors (from Ref.~\cite{Engels:1994xj}) 
    to evaluate the lattice spacing. See text for details.}
    \label{new-renorm}
%  \end{center}
\end{figure}

\begin{figure}[htb!]
%\begin{center}
%  \epsfxsize 80mm
%\hspace{0cm}
\resizebox{3in}{3in}{\includegraphics{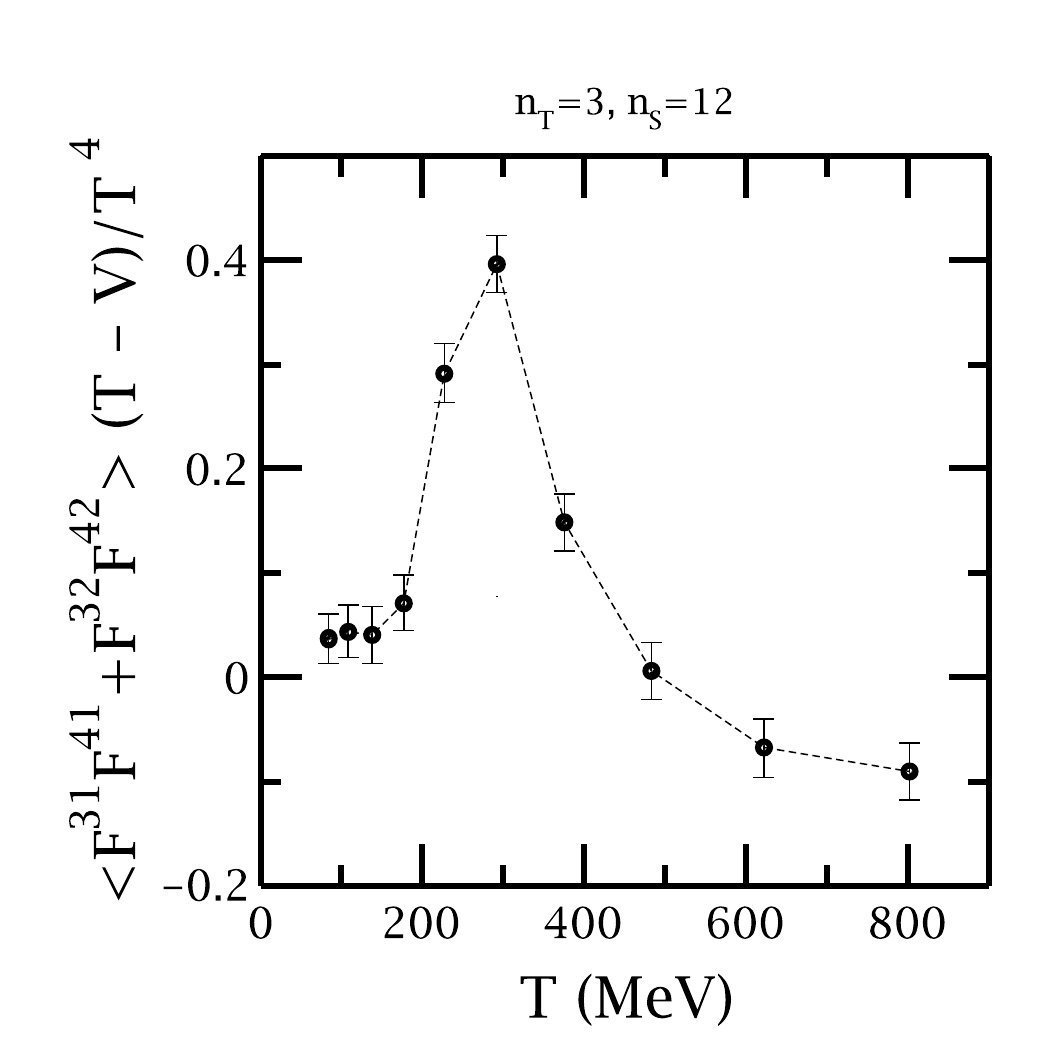}} 
%\vspace{0.25cm}
    \caption{The temperature dependence of the local operator $\lc F^{3x} F^{4x} + F^{3y} F^{4y} \rc$ 
    (thermal contribution minus vacuum contribution) scaled by $T^{4}$ to make it 
    dimensionless. The lattice spacing is set using Eq.~\eqref{lattice-spacing}.}
    \label{fig3}
%  \end{center}
\end{figure}

The formula above, does not provide the best means to set the scale on the lattice at finite temperature~\cite{Engels:1990vr,Engels:1992fs,Fingberg:1992ju}. 
However, it constitutes a simple 
formula that is very easy to use. We have also set the scale using a non-perturbative approach as outlined in Ref.~\cite{Engels:1994xj}
where the formula for the lattice spacing is expressed as the product of that obtained from Eq.~\eqref{lattice-spacing} and a 
non-perturbative function $\lambda(g^{2})$ which has been dialed to ensure that $T_{c}/\Lambda_{L}$ is independent of $g^{2}$, 
%(the value of 
%$\Lambda_{L}$ has been dialed to $10.3$~GeV to maintain the same string tension as in Fig.~\ref{fig2}). 
(comparing with a vacuum string tension of $\sqrt{K}=400$~MeV, this procedure yields a $\Lambda_{L} = 10.3$~MeV).
The results for the field-strength-field-strength correlation $\sum_{i=1,2}( F^{3i} F^{3i} - F^{4i}F^{4i} )/2$ with this next choice of 
formula for the lattice spacing are presented in Fig~\ref{new-renorm}. Again, the correlation results are scaled by $T^{4}$.
In both plots, gauge configurations are generated using a simple 
heat bath algorithm~\cite{Creutz:1984mg}. Calculations consist of 5000 heat bath sweeps for each data point. 
The error represents the standard error as defined in Ref.~\cite{Daniell:1984ea}.

Figures \ref{fig2} and \ref{new-renorm} represent our results for the calculation of the uncrossed operator product 
$\sum_{i=1,2}( F^{31} F^{31} - F^{41}F^{41} )/2$
as a function of 
the temperature, as measured on the lattice. We find that 
while the calculations with $n_{t} = 3,4$ do not show scaling with lattice size, the calculations with $n_{t}=4,5,6$ show 
good scaling especially in Fig.~\ref{new-renorm} where we clearly note the independence of the transition temperature on 
the lattice size. The transition is around $T_{c} \sim 250-350$~MeV for the curves with perturbative renormalization, while 
it is around $T_{c} \sim 150$~MeV for the curves with the non-perturbative factor, with the same choice of $\Lambda_{L}$.
Thus, while the behavior around the transition is  sensitive to the choice of how the scale is set on the lattice, the behavior 
of the correlation at a temperature $T\geq 1.25 - 2 T_{c}$, or in more definite terms 
$T>400$ MeV, seems to be unchanged, i.e. the correlator yields the value of 
$\sim 0.5 T^{4}$.

Given the behavior around the transition temperature along with the larger fluctuations in this region, we will focus on discussing the value of the 
field-strength-field-strength correlator at temperatures above $1.25 - 2 T_{c} $ where the expectation for the correlator has begun to scale with $T^{4}$.
The goal is to evaluate the series of terms outlined in Eq.~\eqref{I1-simple} in this region. 
The plots in Figs.~\ref{fig2},\ref{new-renorm} represent the evaluation 
of a part of the first correlator in this series, as discussed in Eq.~\eqref{no-cross}. The remaining terms are the cross terms $\lc F^{3x} F^{4x} + F^{3y} F^{4y} \rc$, which we have so far neglected. We now present a computation of these terms in Fig.~\ref{fig3} for the case of $n_{t} =3$. The plot 
represents the difference of the finite temperature and vacuum calculations of the same operator product scaled by $T^{4}$, with the lattice 
spacing set by Eq.~\eqref{lattice-spacing}. This plot should be compared with Fig.~\ref{fig2}. While the crossed correlator is of the same size as the 
uncrossed correlator, measured in Fig~\ref{fig2}, in the phase transition region, above a temperature of $T=400$MeV, the crossed correlator is 
rather small compared to the uncrossed correlator $\sum_{i=1,2}( F^{31} F^{31} - F^{41}F^{41} )/2$. 

\begin{figure}[htb!]
%\begin{center}
%  \epsfxsize 80mm
%\hspace{0cm}
\resizebox{3in}{3in}{\includegraphics{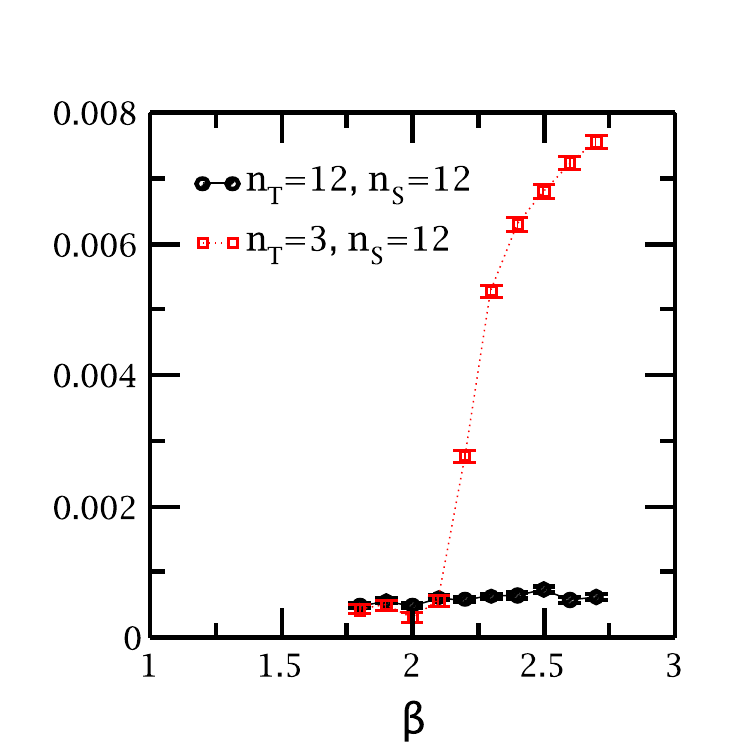}} 
%\vspace{0.25cm}
\caption{(color online) Unscaled expectation of lattice-size-independent correlator 
$\sum_{i=1,2}a_{L}^{4}( F^{31} F^{31} - F^{41}F^{41} )/2$ at finite temperature (red squares) 
versus expectation in vacuum (black circles) as a function of $\beta = 4/g^{2}$($g$ is the bare lattice coupling). 
The plot is for $n_{t} = 3$ and 
$n_{s} = 12$.}
\label{bare_plot_1}
%  \end{center}
\end{figure}

\begin{figure}[htb!]
%\begin{center}
%  \epsfxsize 80mm
%\hspace{0cm}
\resizebox{3in}{3in}{\includegraphics{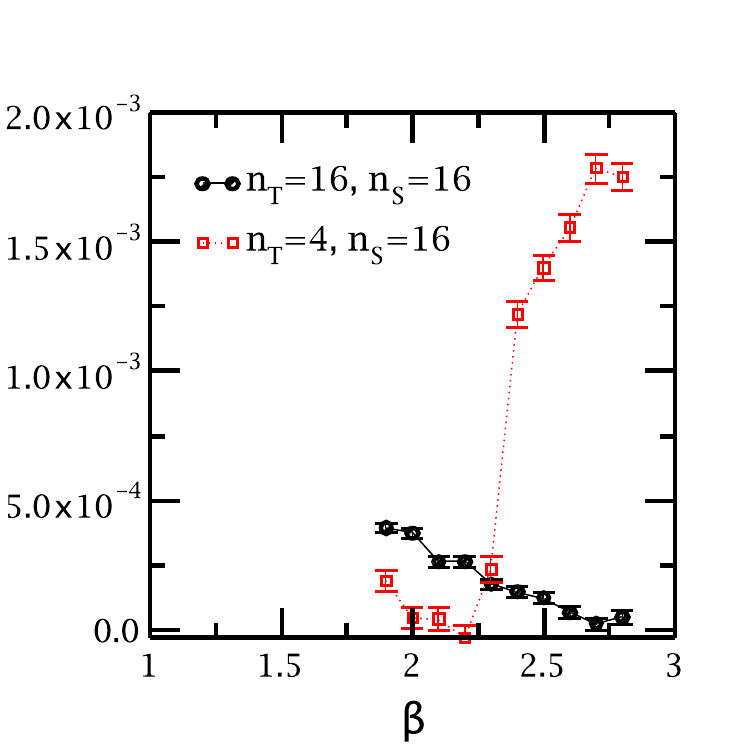}} 
%\vspace{0.25cm}
\caption{(color online) Same as Fig.~\ref{bare_plot_1}, but with $n_{t}=4, n_{s}=16$.}
\label{bare_plot_2}
%  \end{center}
\end{figure}

\begin{figure}[htb!]
%\begin{center}
%  \epsfxsize 80mm
%\hspace{0cm}
\resizebox{3in}{3in}{\includegraphics{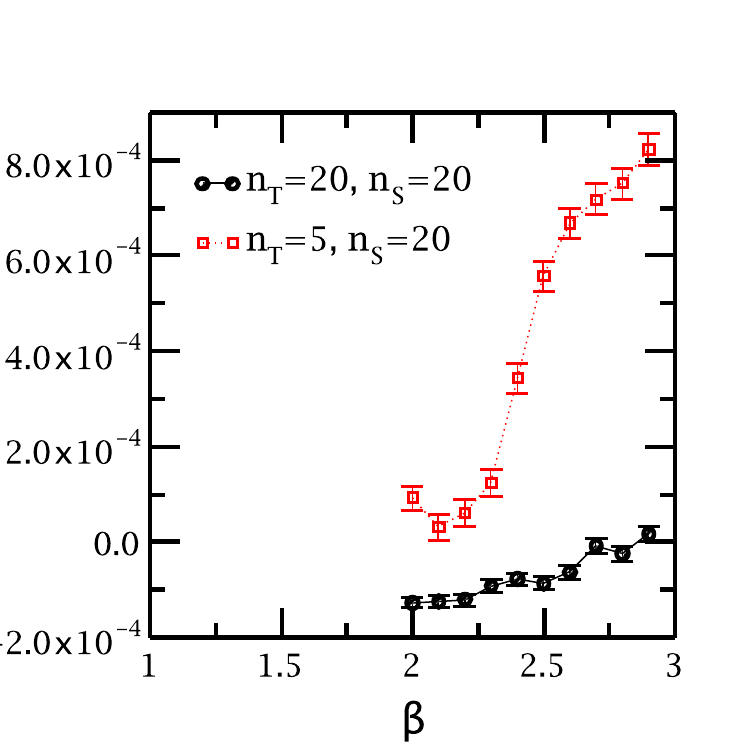}} 
%\vspace{0.25cm}
\caption{(color online) Same as Fig.~\ref{bare_plot_1}, but with $n_{t}=5, n_{s}=20$.}
\label{bare_plot_3}
%  \end{center}
\end{figure}

In the calculation of the cross term, we have considered the difference of the thermal and vacuum expectation values of the operator 
product. For the case of $n_{t} =3$, as presented in Fig.~\ref{fig3}, this is not a very time intensive calculation. 
However, the calculation of the vacuum expectation values 
becomes increasingly numerically intensive with growing $n_{t}$ for both the crossed and uncrossed operator product. 
For the case of $n_{t}=6$, the calculation of the vacuum expectation value has become prohibitively difficult. As a result, in 
Fig.~\ref{fig2} and Fig.~\ref{new-renorm}, only the thermal expectation value of the uncrossed correlator is plotted. This engenders 
a small systematic error as the uncrossed correlators are a difference of two operator products, both of which have vacuum expectation 
values of similar size. As a result, the vacuum expectation values of $\sum_{i=1,2}( F^{3i} F^{3i} - F^{4i}F^{4i} )/2$ are small compared 
to the thermal expectation, particularly in a region far above the transition. 
To illustrate the small size of the vacuum expectation values, we plot the thermal expectation of the uncrossed operator product as a function of 
the bare coupling on the lattice, i.e., without any scaling relation for the lattice spacing. This is plotted for the case of $n_{t}=3$ in Fig.~\ref{bare_plot_1}, 
for the case of $n_{t}=4$ in Fig.~\ref{bare_plot_2}, and for the case of $n_{t}=5$ in Fig.~\ref{bare_plot_3}. As mentioned above, the 
calculation of the expectation of the operator in the vacuum for the case of $n_{t}=6$ has turned out to be prohibitively difficult with current 
resources. As a result, this has not been presented. To plot consistent results, the plots in Figs.~\ref{fig2} and \ref{new-renorm} do not contain any 
vacuum subtraction.

A careful observation of all these curves will indicate that at $T>1.25T_{c}$, the vacuum expectation of the operator product 
$\sum_{i=1,2}( F^{3i} F^{3i} - F^{4i}F^{4i} )/2$ is considerably smaller than the thermal expectation, and so has been ignored in 
the remainder of the discussion. 
We reiterate that, had the focus been on the low temperature region at and below $T_{c}$, one would not be able to 
ignore the vacuum expectation.

\begin{figure}[htbp]
%\begin{center}
%  \epsfxsize 80mm
%\hspace{0cm}
\resizebox{3in}{3in}{\includegraphics{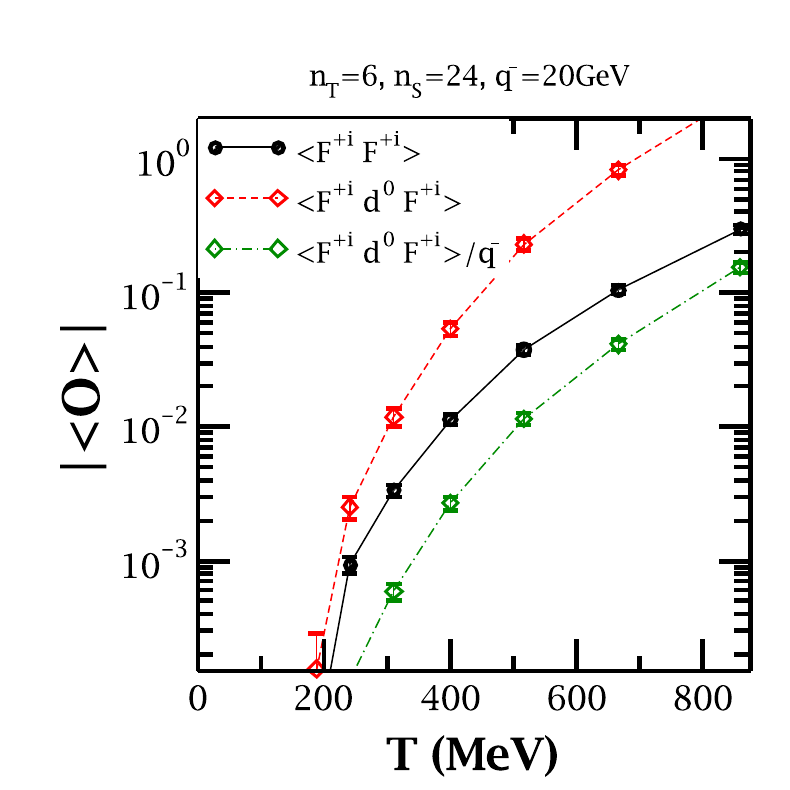}} 
%\vspace{0.25cm}
    \caption{(Color online)The temperature dependence of absolute values of the the local operator $\lc F^{+ i} F^{+ i} \rc$ and the second operator product 
    $\lc \left[ F^{+ i} i \prt^{4} F^{+ i} \right] \rc$, both with (green diamonds) and without (red diamonds) the large factor of $q^{-}$ in the 
    denominator. The lattice spacing is set using Eq.~\eqref{lattice-spacing}. See text for details.}
    \label{fig4}
%  \end{center}
\end{figure}

\begin{figure}[htbp]
%\begin{center}
%  \epsfxsize 80mm
%\hspace{0cm}
\resizebox{3in}{3in}{\includegraphics{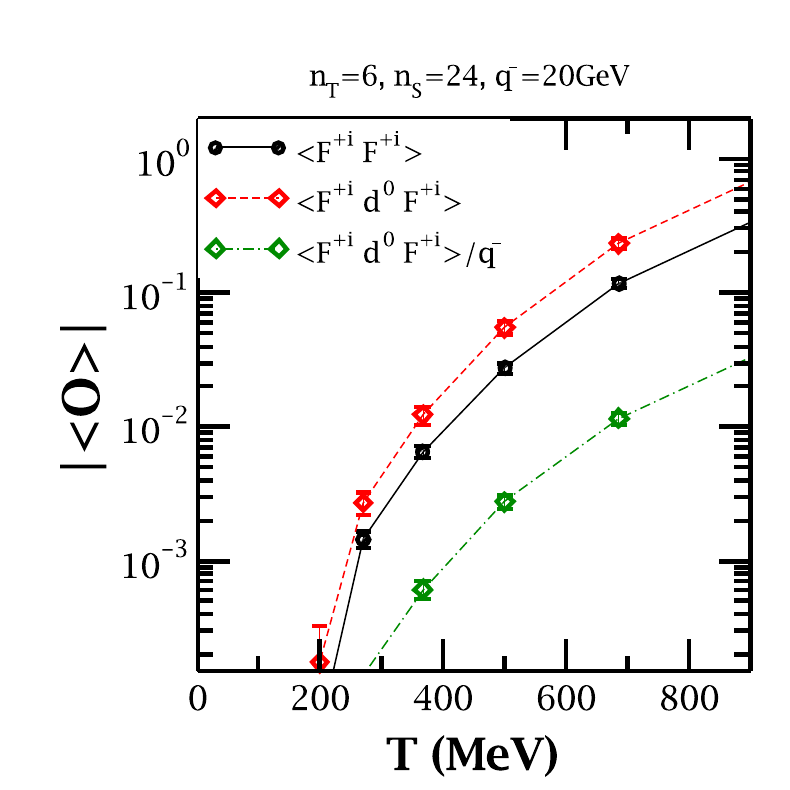}} 
%\vspace{0.25cm}
    \caption{(Color online) Same as Fig.~\ref{fig4}, except for the use of non-perturbative RG factors (from Ref.~\cite{Engels:1994xj}) 
    to evaluate the lattice spacing. See text for details.}
    \label{fig5}
%  \end{center}
\end{figure}

In the preceding paragraphs, we outlined the neglect of a variety of corrections to the leading operator product 
that has to be evaluated to calculate the jet quenching coefficient $\hat{q}$. These corrections tend to be large at 
lower temperatures, at and below $T_{c}$. We will now consider the behavior of the series at higher temperatures.
In our view, the most important correction in this region is brought on by the higher derivative terms in Eqs.~(\ref{I1-Taylor},\ref{I1-simple}). 
To estimate the value of $\hat{q}$ solely from the first term in the expansion in Eq.~\eqref{I1-simple} requires that the 
higher derivative terms be small. 
As an estimate of the size of these terms, we compare the modulus of the expectation of  the first two operators in Eq.~\eqref{I1-simple}, 
for the case of $n_{t}=6$ in Fig.~\ref{fig4}, where the lattice spacing is set using Eq.~\eqref{lattice-spacing}, and in Fig.~\ref{fig5}, where the 
lattice spacing is set using the non-perturbative approach of Ref.~\cite{Engels:1994xj}. 
The next operator in the series is of the form $\sum_{i=1,2} F^{3i} \frac{\mathcal{D}^{4}}{q^{-}} F^{3i} 
- F^{4i} \frac{\mathcal{D}^{4}}{q^{-}} F^{4i} $. We plot it both with (green diamonds) and without (red diamonds) the large factor of $q^{-}$ in 
the denominator. We reiterate again that for this expansion to be a useful estimate of $\hat{q}$, there must be a large jet scale in the problem. 
The results in Fig.~\ref{fig4} are for a $q^{-} = 20$~GeV.

The plots in Figs.~\ref{fig4} and~\ref{fig5} demonstrate that for temperatures below $T=600$~MeV, the expectation of the operator 
$\left[ F^{+ i} i \prt^{4} F^{+ i} \right]/q^{-}$ for a $q^{-} \sim 20$~GeV is less than 25\% of the first operator product (in either method of 
determination of the lattice spacing). It is remarkable that in the more accurate method of determining the lattice spacing, using the 
non-perturbative method of Ref.~\cite{Engels:1994xj}, the leading operator is about a factor of 10 larger than the first correction in the 
vicinity of $T\sim600$~MeV. 
Based on the plots in Figs.~\ref{fig4} and~\ref{fig5}, for temperatures below $600$~MeV, for a $q^{-} \sim 20$~GeV, we may obtain an estimate of the transport 
coefficient $\hat{q}$ from only the leading term in this lattice calculation. 

As pointed out in the prior discussion of 
Figs.~\ref{fig2}-\ref{bare_plot_3}, the corrections from the vacuum expectation of the operator 
product, the uncertainty from scale setting and the larger fluctuation around the transition are small enough only for 
$T>400$~MeV. Thus one can extract $\hat{q}$ 
from such a calculation only in the range $400$~MeV$< T < 600$~MeV. This range coincides with the highest temperatures 
reached at RHIC and LHC and thus will allow future, more sophisticated, efforts to compare meaningfully with the values of $\hat{q}$ obtained from phenomenological analysis of RHIC and LHC data. This constitutes the primary result of the current manuscript: the 
demonstration that the framework developed in Sections III, IV and V can be used to obtain reliable estimates of jet transport 
coefficients in a hot medium. Of course, comparisons with experiment will require both a more sophisticated perturbative 
analysis as well as a much more developed lattice calculation.

\section{Estimating $\hat{q}$ and concluding discussions}

In this concluding section, we attempt a simple minded extraction of the jet transport coefficient $\hat{q}$ from the lattice calculation outlined above. 
We would like to clearly point out that what follows is, for the most part, a hand waving estimate. Later calculations, which will involve the 
parton being produced far off its mass shell and radiating gluons as it propagates, will involve many more issues in the extraction of $\hat{q}$.
Recall that our calculation required that the hard quark moves through the medium without undergoing any radiation.
This constrains the highest virtuality that the quark may possess for such an approach to make sense. In a future effort, 
partons with a higher initial virtuality will be considered. These will undergo radiative splitting in the medium and may show 
sensitivity to a somewhat different series of operator products.

We choose the region around the 3rd last point in the $\lc F^{+ i} F^{+ i} \rc $ 
curve in Fig.~\ref{fig2} and Fig.~\ref{fig4}. This corresponds to a temperature of $T \simeq 400$~MeV, 
which is in between the top temperature reached at RHIC and LHC collisions. At a $T\simeq 400$~MeV, $\lc F^{+ i} F^{+ i}\rc = 0.01$~GeV$^{4}$.
Also we are considering a lattice with a length given by $4 \times n_{t} a_{L} = 4/0.4$~GeV$^{-1} = 10$~GeV$^{-1}$. 
This states that the maximum virtuality of a jet (with a $q^{-} = 20$~GeV) which traverses such a length without undergoing 
radiation is given as $\mu^{2} = E/L = 20/10/\sqrt{2} \simeq 1.4$~GeV$^{2}$. Thus $Q^{+}=1.4/40$~GeV. With these estimates, 
we obtain, 
\bea
\hat{\bar{q}} = \frac{2 \sqrt{2} \pi^{2} \A_{s}(\mu^{2})  }{N_{c} 2 Q^{+} q^{-} }  \lc M | F^{+ i}  F^{+ i}  | M \rc.
\eea
Using $\A_{s} (1.4 \rm{GeV}^{2}) = 0.375$~\cite{Kluth:2007np}, we obtain $\hat{\bar{q}} = 0.186 \rm{GeV}^{2}/{\rm fm}$ for an 
$SU(2)$ quark traversing a quenched $SU(2)$ plasma. In most phenomenological estimates one quotes the $\hat{q}$ of the 
gluon. 
If the above calculation were done for an $SU(2)$ gluon, the $\hat{q}$ would 
differ only by the overall Casimir factor of $C_{A}/C_{F} = 2N_{c}^{2}/(N_{c}^{2} - 1) = 8/3$ yielding a $\hat{q}_{G} = 0.5$~GeV$^{2}$/fm, 
at a $T=400$~MeV.

In future efforts, the calculation will be extended to higher statistics runs, along with a more careful treatment including the crossed correlators, 
to evaluate the $\hat{q}$ across the phase 
transition. The next step is to evaluate the required operator products for a realistic jet which starts at a higher 
virtuality and undergoes radiative splitting in the medium. In such a calculation, the range of operators that will 
need to be evaluated [i.e., number of terms in the series of Eq.~\eqref{I1-simple} that need to be retained] 
will depend on the particular parton in the shower, in particular on that parton's energy and virtuality.

Finally, to be of use to jet modification at RHIC and LHC, the calculation will have to be extended to 
unquenched $SU(3)$. This will involve a non-trivial extension, not only due to the increase in the level of 
computational difficulty, but also due to the issues arising from the larger gauge group. 
Beyond these extensions, more sophisticated renormalization factors will have to be introduced, and better means to set the lattice 
spacing will have to be used. 
At this stage we may only set suggestive limits on such a future estimation: 
the quenched $SU(2)$ calculation has $3$ colors of gluons as the fundamental fields in its Lagrangian, whereas there 
are $8$ colors of gluons in quenched $SU(3)$, along with $3$ colors of quarks and antiquarks in the unquenched 
$SU(3)$ calculation (Note that even if the plasma were completely perturbative, quarks would contribute differently 
to the calculation of $\hat{q}$ than gluons~\cite{Majumder:2007zh}, or rather the lattice calculation could change considerably with 
the introduction of dynamical fermions). 
Ignoring such subtleties, assuming 2 flavors of light quarks, 
and assuming a simple scaling law with number of fields in the Lagrangian, we estimate that the 
full $\hat{q}$ at RHIC would lie in the range:
\bea
\mbx\!\!\!\hat{q} (T = 400 {\rm MeV}) = 1.3 {\rm GeV}^{2}/{\rm fm} - 3.3  {\rm GeV}^{2}/{\rm fm}.
\eea
(If we had instead used Figs.~\ref{new-renorm} and \ref{fig5}, to estimate $\hat{q}$ we would have obtained a range 
from $0.9 - 2.3$~GeV$^{2}$/fm.)
We should point out that while the above estimate is very specific to a particular range of $q^{+}, q^{-}$ of the 
propagating parton, the estimate obtained form phenomenological analysis of RHIC collisions is an average over 
a wide range of parton energies and virtualities. In spite of the many shortcomings of the above calculation, 
we find the very encouraging result that our estimate for 
$\hat{q}$ at $T=400$~MeV is comparable with that extracted from phenomenological analysis of RHIC and 
LHC data~\cite{Bass:2008rv,Majumder:2011uk}. 

\acknowledgements
The author would like to thank J.~Drut, C.~Gale, S.~Gavin, U.~Heinz, Y.~Kovchegov, B.~M\"{u}ller, A.~Sch\"{a}fer and J.~Shigemitsu for 
helpful discussions. This work was supported in part by the National Science Foundation under grant number PHY-1207918. 
Part of this work was carried out while the author was employed at Ohio State University, where it 
was supported in part by the U.S. Department of Energy under grant no. DE-SC0004286 and (within the framework of the JET 
collaboration) DE-SC0004104.

\bibliography{refs}

\begin{thebibliography}{46}
\expandafter\ifx\csname natexlab\endcsname\relax\def\natexlab#1{#1}\fi
\expandafter\ifx\csname bibnamefont\endcsname\relax
  \def\bibnamefont#1{#1}\fi
\expandafter\ifx\csname bibfnamefont\endcsname\relax
  \def\bibfnamefont#1{#1}\fi
\expandafter\ifx\csname citenamefont\endcsname\relax
  \def\citenamefont#1{#1}\fi
\expandafter\ifx\csname url\endcsname\relax
  \def\url#1{\texttt{#1}}\fi
\expandafter\ifx\csname urlprefix\endcsname\relax\def\urlprefix{URL }\fi
\providecommand{\bibinfo}[2]{#2}
\providecommand{\eprint}[2][]{\url{#2}}

\bibitem[{\citenamefont{Adcox et~al.}(2005)}]{Adcox:2004mh}
\bibinfo{author}{\bibfnamefont{K.}~\bibnamefont{Adcox}} \bibnamefont{et~al.}
  (\bibinfo{collaboration}{PHENIX}), \bibinfo{journal}{Nucl. Phys.}
  \textbf{\bibinfo{volume}{A757}}, \bibinfo{pages}{184} (\bibinfo{year}{2005}),
  \eprint{nucl-ex/0410003}.

\bibitem[{\citenamefont{Adams et~al.}(2005)}]{Adams:2005dq}
\bibinfo{author}{\bibfnamefont{J.}~\bibnamefont{Adams}} \bibnamefont{et~al.}
  (\bibinfo{collaboration}{STAR}), \bibinfo{journal}{Nucl. Phys.}
  \textbf{\bibinfo{volume}{A757}}, \bibinfo{pages}{102} (\bibinfo{year}{2005}),
  \eprint{nucl-ex/0501009}.

\bibitem[{\citenamefont{Schukraft}(2011)}]{Schukraft:2011ch}
\bibinfo{author}{\bibfnamefont{J.}~\bibnamefont{Schukraft}}
  (\bibinfo{year}{2011}), \eprint{1112.0550}.

\bibitem[{\citenamefont{Steinberg and Collaboration}(2011)}]{Steinberg:2011qq}
\bibinfo{author}{\bibfnamefont{P.}~\bibnamefont{Steinberg}} \bibnamefont{and}
  \bibinfo{author}{\bibfnamefont{A.}~\bibnamefont{Collaboration}}
  (\bibinfo{year}{2011}), \eprint{1110.3352}.

\bibitem[{\citenamefont{Wyslouch}(2011)}]{Wyslouch:2011zz}
\bibinfo{author}{\bibfnamefont{B.}~\bibnamefont{Wyslouch}}
  (\bibinfo{collaboration}{CMS Collaboration}), \bibinfo{journal}{J.Phys.G}
  \textbf{\bibinfo{volume}{G38}}, \bibinfo{pages}{124005}
  (\bibinfo{year}{2011}).

\bibitem[{\citenamefont{Armesto et~al.}(2004)\citenamefont{Armesto, Salgado,
  and Wiedemann}}]{Armesto:2003jh}
\bibinfo{author}{\bibfnamefont{N.}~\bibnamefont{Armesto}},
  \bibinfo{author}{\bibfnamefont{C.~A.} \bibnamefont{Salgado}},
  \bibnamefont{and} \bibinfo{author}{\bibfnamefont{U.~A.}
  \bibnamefont{Wiedemann}}, \bibinfo{journal}{Phys. Rev.}
  \textbf{\bibinfo{volume}{D69}}, \bibinfo{pages}{114003}
  (\bibinfo{year}{2004}), \eprint{hep-ph/0312106}.

\bibitem[{\citenamefont{Salgado and Wiedemann}(2003)}]{Salgado:2003gb}
\bibinfo{author}{\bibfnamefont{C.~A.} \bibnamefont{Salgado}} \bibnamefont{and}
  \bibinfo{author}{\bibfnamefont{U.~A.} \bibnamefont{Wiedemann}},
  \bibinfo{journal}{Phys. Rev.} \textbf{\bibinfo{volume}{D68}},
  \bibinfo{pages}{014008} (\bibinfo{year}{2003}), \eprint{hep-ph/0302184}.

\bibitem[{\citenamefont{Wiedemann}(2000)}]{Wiedemann:2000za}
\bibinfo{author}{\bibfnamefont{U.~A.} \bibnamefont{Wiedemann}},
  \bibinfo{journal}{Nucl. Phys.} \textbf{\bibinfo{volume}{B588}},
  \bibinfo{pages}{303} (\bibinfo{year}{2000}), \eprint{hep-ph/0005129}.

\bibitem[{\citenamefont{Arnold et~al.}(2002)\citenamefont{Arnold, Moore, and
  Yaffe}}]{Arnold:2002ja}
\bibinfo{author}{\bibfnamefont{P.}~\bibnamefont{Arnold}},
  \bibinfo{author}{\bibfnamefont{G.~D.} \bibnamefont{Moore}}, \bibnamefont{and}
  \bibinfo{author}{\bibfnamefont{L.~G.} \bibnamefont{Yaffe}},
  \bibinfo{journal}{JHEP} \textbf{\bibinfo{volume}{06}}, \bibinfo{pages}{030}
  (\bibinfo{year}{2002}), \eprint{hep-ph/0204343}.

\bibitem[{\citenamefont{Jeon and Moore}(2005)}]{Jeon:2003gi}
\bibinfo{author}{\bibfnamefont{S.}~\bibnamefont{Jeon}} \bibnamefont{and}
  \bibinfo{author}{\bibfnamefont{G.~D.} \bibnamefont{Moore}},
  \bibinfo{journal}{Phys. Rev.} \textbf{\bibinfo{volume}{C71}},
  \bibinfo{pages}{034901} (\bibinfo{year}{2005}), \eprint{hep-ph/0309332}.

\bibitem[{\citenamefont{Qin et~al.}(2009)\citenamefont{Qin, Ruppert, Gale,
  Jeon, and Moore}}]{Qin:2009bk}
\bibinfo{author}{\bibfnamefont{G.-Y.} \bibnamefont{Qin}},
  \bibinfo{author}{\bibfnamefont{J.}~\bibnamefont{Ruppert}},
  \bibinfo{author}{\bibfnamefont{C.}~\bibnamefont{Gale}},
  \bibinfo{author}{\bibfnamefont{S.}~\bibnamefont{Jeon}}, \bibnamefont{and}
  \bibinfo{author}{\bibfnamefont{G.~D.} \bibnamefont{Moore}}
  (\bibinfo{year}{2009}), \eprint{0906.3280}.

\bibitem[{\citenamefont{Gyulassy et~al.}(2001)\citenamefont{Gyulassy, Levai,
  and Vitev}}]{Gyulassy:2000er}
\bibinfo{author}{\bibfnamefont{M.}~\bibnamefont{Gyulassy}},
  \bibinfo{author}{\bibfnamefont{P.}~\bibnamefont{Levai}}, \bibnamefont{and}
  \bibinfo{author}{\bibfnamefont{I.}~\bibnamefont{Vitev}},
  \bibinfo{journal}{Nucl. Phys.} \textbf{\bibinfo{volume}{B594}},
  \bibinfo{pages}{371} (\bibinfo{year}{2001}), \eprint{nucl-th/0006010}.

\bibitem[{\citenamefont{Gyulassy et~al.}(2000)\citenamefont{Gyulassy, Levai,
  and Vitev}}]{Gyulassy:2000fs}
\bibinfo{author}{\bibfnamefont{M.}~\bibnamefont{Gyulassy}},
  \bibinfo{author}{\bibfnamefont{P.}~\bibnamefont{Levai}}, \bibnamefont{and}
  \bibinfo{author}{\bibfnamefont{I.}~\bibnamefont{Vitev}},
  \bibinfo{journal}{Phys. Rev. Lett.} \textbf{\bibinfo{volume}{85}},
  \bibinfo{pages}{5535} (\bibinfo{year}{2000}), \eprint{nucl-th/0005032}.

\bibitem[{\citenamefont{Gyulassy et~al.}(2002)\citenamefont{Gyulassy, Levai,
  and Vitev}}]{Gyulassy:2001nm}
\bibinfo{author}{\bibfnamefont{M.}~\bibnamefont{Gyulassy}},
  \bibinfo{author}{\bibfnamefont{P.}~\bibnamefont{Levai}}, \bibnamefont{and}
  \bibinfo{author}{\bibfnamefont{I.}~\bibnamefont{Vitev}},
  \bibinfo{journal}{Phys. Lett.} \textbf{\bibinfo{volume}{B538}},
  \bibinfo{pages}{282} (\bibinfo{year}{2002}), \eprint{nucl-th/0112071}.

\bibitem[{\citenamefont{Guo and Wang}(2000)}]{Guo:2000nz}
\bibinfo{author}{\bibfnamefont{X.-F.} \bibnamefont{Guo}} \bibnamefont{and}
  \bibinfo{author}{\bibfnamefont{X.-N.} \bibnamefont{Wang}},
  \bibinfo{journal}{Phys. Rev. Lett.} \textbf{\bibinfo{volume}{85}},
  \bibinfo{pages}{3591} (\bibinfo{year}{2000}), \eprint{hep-ph/0005044}.

\bibitem[{\citenamefont{Wang and Guo}(2001)}]{Wang:2001ifa}
\bibinfo{author}{\bibfnamefont{X.-N.} \bibnamefont{Wang}} \bibnamefont{and}
  \bibinfo{author}{\bibfnamefont{X.-F.} \bibnamefont{Guo}},
  \bibinfo{journal}{Nucl. Phys.} \textbf{\bibinfo{volume}{A696}},
  \bibinfo{pages}{788} (\bibinfo{year}{2001}), \eprint{hep-ph/0102230}.

\bibitem[{\citenamefont{Majumder and Muller}(2008)}]{Majumder:2007hx}
\bibinfo{author}{\bibfnamefont{A.}~\bibnamefont{Majumder}} \bibnamefont{and}
  \bibinfo{author}{\bibfnamefont{B.}~\bibnamefont{Muller}},
  \bibinfo{journal}{Phys. Rev.} \textbf{\bibinfo{volume}{C77}},
  \bibinfo{pages}{054903} (\bibinfo{year}{2008}), \eprint{0705.1147}.

\bibitem[{\citenamefont{Majumder et~al.}(2008)\citenamefont{Majumder, Fries,
  and Muller}}]{Majumder:2007ne}
\bibinfo{author}{\bibfnamefont{A.}~\bibnamefont{Majumder}},
  \bibinfo{author}{\bibfnamefont{R.~J.} \bibnamefont{Fries}}, \bibnamefont{and}
  \bibinfo{author}{\bibfnamefont{B.}~\bibnamefont{Muller}},
  \bibinfo{journal}{Phys. Rev.} \textbf{\bibinfo{volume}{C77}},
  \bibinfo{pages}{065209} (\bibinfo{year}{2008}), \eprint{0711.2475}.

\bibitem[{\citenamefont{Majumder}(2009)}]{Majumder:2009ge}
\bibinfo{author}{\bibfnamefont{A.}~\bibnamefont{Majumder}}
  (\bibinfo{year}{2009}), \eprint{0912.2987}.

\bibitem[{\citenamefont{Gubser}(2006)}]{Gubser:2006bz}
\bibinfo{author}{\bibfnamefont{S.~S.} \bibnamefont{Gubser}},
  \bibinfo{journal}{Phys. Rev.} \textbf{\bibinfo{volume}{D74}},
  \bibinfo{pages}{126005} (\bibinfo{year}{2006}), \eprint{hep-th/0605182}.

\bibitem[{\citenamefont{Gubser et~al.}(2009)\citenamefont{Gubser, Pufu, Rocha,
  and Yarom}}]{Gubser:2009sn}
\bibinfo{author}{\bibfnamefont{S.~S.} \bibnamefont{Gubser}},
  \bibinfo{author}{\bibfnamefont{S.~S.} \bibnamefont{Pufu}},
  \bibinfo{author}{\bibfnamefont{F.~D.} \bibnamefont{Rocha}}, \bibnamefont{and}
  \bibinfo{author}{\bibfnamefont{A.}~\bibnamefont{Yarom}}
  (\bibinfo{year}{2009}), \eprint{0902.4041}.

\bibitem[{\citenamefont{Herzog et~al.}(2006)\citenamefont{Herzog, Karch,
  Kovtun, Kozcaz, and Yaffe}}]{Herzog:2006gh}
\bibinfo{author}{\bibfnamefont{C.~P.} \bibnamefont{Herzog}},
  \bibinfo{author}{\bibfnamefont{A.}~\bibnamefont{Karch}},
  \bibinfo{author}{\bibfnamefont{P.}~\bibnamefont{Kovtun}},
  \bibinfo{author}{\bibfnamefont{C.}~\bibnamefont{Kozcaz}}, \bibnamefont{and}
  \bibinfo{author}{\bibfnamefont{L.~G.} \bibnamefont{Yaffe}},
  \bibinfo{journal}{JHEP} \textbf{\bibinfo{volume}{07}}, \bibinfo{pages}{013}
  (\bibinfo{year}{2006}), \eprint{hep-th/0605158}.

\bibitem[{\citenamefont{Casalderrey-Solana and
  Teaney}(2007)}]{CasalderreySolana:2007qw}
\bibinfo{author}{\bibfnamefont{J.}~\bibnamefont{Casalderrey-Solana}}
  \bibnamefont{and} \bibinfo{author}{\bibfnamefont{D.}~\bibnamefont{Teaney}},
  \bibinfo{journal}{JHEP} \textbf{\bibinfo{volume}{0704}}, \bibinfo{pages}{039}
  (\bibinfo{year}{2007}), \eprint{hep-th/0701123}.

\bibitem[{\citenamefont{Majumder and Van~Leeuwen}(2011)}]{Majumder:2010qh}
\bibinfo{author}{\bibfnamefont{A.}~\bibnamefont{Majumder}} \bibnamefont{and}
  \bibinfo{author}{\bibfnamefont{M.}~\bibnamefont{Van~Leeuwen}},
  \bibinfo{journal}{Prog.Part.Nucl.Phys.} \textbf{\bibinfo{volume}{A66}},
  \bibinfo{pages}{41} (\bibinfo{year}{2011}), \eprint{1002.2206}.

\bibitem[{\citenamefont{Armesto et~al.}(2012)\citenamefont{Armesto, Cole, Gale,
  Horowitz, Jacobs et~al.}}]{Armesto:2011ht}
\bibinfo{author}{\bibfnamefont{N.}~\bibnamefont{Armesto}},
  \bibinfo{author}{\bibfnamefont{B.}~\bibnamefont{Cole}},
  \bibinfo{author}{\bibfnamefont{C.}~\bibnamefont{Gale}},
  \bibinfo{author}{\bibfnamefont{W.~A.} \bibnamefont{Horowitz}},
  \bibinfo{author}{\bibfnamefont{P.}~\bibnamefont{Jacobs}},
  \bibnamefont{et~al.}, \bibinfo{journal}{Phys.Rev.}
  \textbf{\bibinfo{volume}{C86}}, \bibinfo{pages}{064904}
  (\bibinfo{year}{2012}), \eprint{1106.1106}.

\bibitem[{\citenamefont{Braaten and
  Pisarski}(1990{\natexlab{a}})}]{Braaten:1989kk}
\bibinfo{author}{\bibfnamefont{E.}~\bibnamefont{Braaten}} \bibnamefont{and}
  \bibinfo{author}{\bibfnamefont{R.~D.} \bibnamefont{Pisarski}},
  \bibinfo{journal}{Phys. Rev. Lett.} \textbf{\bibinfo{volume}{64}},
  \bibinfo{pages}{1338} (\bibinfo{year}{1990}{\natexlab{a}}).

\bibitem[{\citenamefont{Braaten and
  Pisarski}(1990{\natexlab{b}})}]{Braaten:1989mz}
\bibinfo{author}{\bibfnamefont{E.}~\bibnamefont{Braaten}} \bibnamefont{and}
  \bibinfo{author}{\bibfnamefont{R.~D.} \bibnamefont{Pisarski}},
  \bibinfo{journal}{Nucl.Phys.} \textbf{\bibinfo{volume}{B337}},
  \bibinfo{pages}{569} (\bibinfo{year}{1990}{\natexlab{b}}).

\bibitem[{\citenamefont{Horowitz}(2011)}]{Horowitz:2011wm}
\bibinfo{author}{\bibfnamefont{W.}~\bibnamefont{Horowitz}}
  (\bibinfo{year}{2011}), \eprint{1108.5876}.

\bibitem[{\citenamefont{Idilbi and Majumder}(2009)}]{Idilbi:2008vm}
\bibinfo{author}{\bibfnamefont{A.}~\bibnamefont{Idilbi}} \bibnamefont{and}
  \bibinfo{author}{\bibfnamefont{A.}~\bibnamefont{Majumder}},
  \bibinfo{journal}{Phys.Rev.} \textbf{\bibinfo{volume}{D80}},
  \bibinfo{pages}{054022} (\bibinfo{year}{2009}), \eprint{0808.1087}.

\bibitem[{\citenamefont{Garcia-Echevarria
  et~al.}(2011)\citenamefont{Garcia-Echevarria, Idilbi, and
  Scimemi}}]{GarciaEchevarria:2011md}
\bibinfo{author}{\bibfnamefont{M.}~\bibnamefont{Garcia-Echevarria}},
  \bibinfo{author}{\bibfnamefont{A.}~\bibnamefont{Idilbi}}, \bibnamefont{and}
  \bibinfo{author}{\bibfnamefont{I.}~\bibnamefont{Scimemi}},
  \bibinfo{journal}{Phys. Rev.} \textbf{\bibinfo{volume}{D84}},
  \bibinfo{pages}{011502} (\bibinfo{year}{2011}).

\bibitem[{\citenamefont{Kronfeld and Photiadis}(1985)}]{Kronfeld:1984zv}
\bibinfo{author}{\bibfnamefont{A.~S.} \bibnamefont{Kronfeld}} \bibnamefont{and}
  \bibinfo{author}{\bibfnamefont{D.~M.} \bibnamefont{Photiadis}},
  \bibinfo{journal}{Phys.Rev.} \textbf{\bibinfo{volume}{D31}},
  \bibinfo{pages}{2939} (\bibinfo{year}{1985}).

\bibitem[{\citenamefont{Martinelli and Sachrajda}(1987)}]{Martinelli:1987zd}
\bibinfo{author}{\bibfnamefont{G.}~\bibnamefont{Martinelli}} \bibnamefont{and}
  \bibinfo{author}{\bibfnamefont{C.~T.} \bibnamefont{Sachrajda}},
  \bibinfo{journal}{Phys.Lett.} \textbf{\bibinfo{volume}{B196}},
  \bibinfo{pages}{184} (\bibinfo{year}{1987}).

\bibitem[{\citenamefont{Martinelli and
  Sachrajda}(1989{\natexlab{a}})}]{Martinelli:1988rr}
\bibinfo{author}{\bibfnamefont{G.}~\bibnamefont{Martinelli}} \bibnamefont{and}
  \bibinfo{author}{\bibfnamefont{C.~T.} \bibnamefont{Sachrajda}},
  \bibinfo{journal}{Nucl.Phys.} \textbf{\bibinfo{volume}{B316}},
  \bibinfo{pages}{355} (\bibinfo{year}{1989}{\natexlab{a}}).

\bibitem[{\citenamefont{Martinelli and
  Sachrajda}(1989{\natexlab{b}})}]{Martinelli:1988xs}
\bibinfo{author}{\bibfnamefont{G.}~\bibnamefont{Martinelli}} \bibnamefont{and}
  \bibinfo{author}{\bibfnamefont{C.~T.} \bibnamefont{Sachrajda}},
  \bibinfo{journal}{Phys.Lett.} \textbf{\bibinfo{volume}{B217}},
  \bibinfo{pages}{319} (\bibinfo{year}{1989}{\natexlab{b}}).

\bibitem[{\citenamefont{Majumder and Wang}(2004)}]{Majumder:2004wh}
\bibinfo{author}{\bibfnamefont{A.}~\bibnamefont{Majumder}} \bibnamefont{and}
  \bibinfo{author}{\bibfnamefont{X.-N.} \bibnamefont{Wang}},
  \bibinfo{journal}{Phys. Rev.} \textbf{\bibinfo{volume}{D70}},
  \bibinfo{pages}{014007} (\bibinfo{year}{2004}), \eprint{hep-ph/0402245}.

\bibitem[{\citenamefont{Majumder and Shen}(2012)}]{Majumder:2011uk}
\bibinfo{author}{\bibfnamefont{A.}~\bibnamefont{Majumder}} \bibnamefont{and}
  \bibinfo{author}{\bibfnamefont{C.}~\bibnamefont{Shen}},
  \bibinfo{journal}{Phys.Rev.Lett.} \textbf{\bibinfo{volume}{109}},
  \bibinfo{pages}{202301} (\bibinfo{year}{2012}), \eprint{1103.0809}.

\bibitem[{\citenamefont{Engels et~al.}(1981)\citenamefont{Engels, Karsch, Satz,
  and Montvay}}]{Engels:1980ty}
\bibinfo{author}{\bibfnamefont{J.}~\bibnamefont{Engels}},
  \bibinfo{author}{\bibfnamefont{F.}~\bibnamefont{Karsch}},
  \bibinfo{author}{\bibfnamefont{H.}~\bibnamefont{Satz}}, \bibnamefont{and}
  \bibinfo{author}{\bibfnamefont{I.}~\bibnamefont{Montvay}},
  \bibinfo{journal}{Phys.Lett.} \textbf{\bibinfo{volume}{B101}},
  \bibinfo{pages}{89} (\bibinfo{year}{1981}).

\bibitem[{\citenamefont{Creutz}(1984)}]{Creutz:1984mg}
\bibinfo{author}{\bibfnamefont{M.}~\bibnamefont{Creutz}},
  \emph{\bibinfo{title}{{Quarks, Gluons and Lattices}}}
  (\bibinfo{publisher}{Cambridge University Press}, \bibinfo{year}{1984}).

\bibitem[{\citenamefont{Engels et~al.}(1990)\citenamefont{Engels, Fingberg,
  Karsch, Miller, and Weber}}]{Engels:1990vr}
\bibinfo{author}{\bibfnamefont{J.}~\bibnamefont{Engels}},
  \bibinfo{author}{\bibfnamefont{J.}~\bibnamefont{Fingberg}},
  \bibinfo{author}{\bibfnamefont{F.}~\bibnamefont{Karsch}},
  \bibinfo{author}{\bibfnamefont{D.}~\bibnamefont{Miller}}, \bibnamefont{and}
  \bibinfo{author}{\bibfnamefont{M.}~\bibnamefont{Weber}},
  \bibinfo{journal}{Phys.Lett.} \textbf{\bibinfo{volume}{B252}},
  \bibinfo{pages}{625} (\bibinfo{year}{1990}).

\bibitem[{\citenamefont{Engels et~al.}(1992)\citenamefont{Engels, Fingberg, and
  Miller}}]{Engels:1992fs}
\bibinfo{author}{\bibfnamefont{J.}~\bibnamefont{Engels}},
  \bibinfo{author}{\bibfnamefont{J.}~\bibnamefont{Fingberg}}, \bibnamefont{and}
  \bibinfo{author}{\bibfnamefont{D.}~\bibnamefont{Miller}},
  \bibinfo{journal}{Nucl.Phys.} \textbf{\bibinfo{volume}{B387}},
  \bibinfo{pages}{501} (\bibinfo{year}{1992}).

\bibitem[{\citenamefont{Fingberg et~al.}(1993)\citenamefont{Fingberg, Heller,
  and Karsch}}]{Fingberg:1992ju}
\bibinfo{author}{\bibfnamefont{J.}~\bibnamefont{Fingberg}},
  \bibinfo{author}{\bibfnamefont{U.~M.} \bibnamefont{Heller}},
  \bibnamefont{and} \bibinfo{author}{\bibfnamefont{F.}~\bibnamefont{Karsch}},
  \bibinfo{journal}{Nucl.Phys.} \textbf{\bibinfo{volume}{B392}},
  \bibinfo{pages}{493} (\bibinfo{year}{1993}), \eprint{hep-lat/9208012}.

\bibitem[{\citenamefont{Engels et~al.}(1995)\citenamefont{Engels, Karsch, and
  Redlich}}]{Engels:1994xj}
\bibinfo{author}{\bibfnamefont{J.}~\bibnamefont{Engels}},
  \bibinfo{author}{\bibfnamefont{F.}~\bibnamefont{Karsch}}, \bibnamefont{and}
  \bibinfo{author}{\bibfnamefont{K.}~\bibnamefont{Redlich}},
  \bibinfo{journal}{Nucl.Phys.} \textbf{\bibinfo{volume}{B435}},
  \bibinfo{pages}{295} (\bibinfo{year}{1995}), \eprint{hep-lat/9408009}.

\bibitem[{\citenamefont{Daniell et~al.}(1984)\citenamefont{Daniell, Hey, and
  Mandula}}]{Daniell:1984ea}
\bibinfo{author}{\bibfnamefont{G.}~\bibnamefont{Daniell}},
  \bibinfo{author}{\bibfnamefont{A.}~\bibnamefont{Hey}}, \bibnamefont{and}
  \bibinfo{author}{\bibfnamefont{J.}~\bibnamefont{Mandula}},
  \bibinfo{journal}{Phys.Rev.} \textbf{\bibinfo{volume}{D30}},
  \bibinfo{pages}{2230} (\bibinfo{year}{1984}).

\bibitem[{\citenamefont{Kluth}(2008)}]{Kluth:2007np}
\bibinfo{author}{\bibfnamefont{S.}~\bibnamefont{Kluth}},
  \bibinfo{journal}{J.Phys.Conf.Ser.} \textbf{\bibinfo{volume}{110}},
  \bibinfo{pages}{022023} (\bibinfo{year}{2008}), \eprint{0709.0173}.

\bibitem[{\citenamefont{Majumder et~al.}(2007)\citenamefont{Majumder, Muller,
  and Wang}}]{Majumder:2007zh}
\bibinfo{author}{\bibfnamefont{A.}~\bibnamefont{Majumder}},
  \bibinfo{author}{\bibfnamefont{B.}~\bibnamefont{Muller}}, \bibnamefont{and}
  \bibinfo{author}{\bibfnamefont{X.-N.} \bibnamefont{Wang}},
  \bibinfo{journal}{Phys. Rev. Lett.} \textbf{\bibinfo{volume}{99}},
  \bibinfo{pages}{192301} (\bibinfo{year}{2007}), \eprint{hep-ph/0703082}.

\bibitem[{\citenamefont{Bass et~al.}(2009)\citenamefont{Bass, Gale, Majumder,
  Nonaka, Qin et~al.}}]{Bass:2008rv}
\bibinfo{author}{\bibfnamefont{S.~A.} \bibnamefont{Bass}},
  \bibinfo{author}{\bibfnamefont{C.}~\bibnamefont{Gale}},
  \bibinfo{author}{\bibfnamefont{A.}~\bibnamefont{Majumder}},
  \bibinfo{author}{\bibfnamefont{C.}~\bibnamefont{Nonaka}},
  \bibinfo{author}{\bibfnamefont{G.-Y.} \bibnamefont{Qin}},
  \bibnamefont{et~al.}, \bibinfo{journal}{Phys.Rev.}
  \textbf{\bibinfo{volume}{C79}}, \bibinfo{pages}{024901}
  (\bibinfo{year}{2009}), \eprint{0808.0908}.

\end{thebibliography}

\end{document}